\documentclass[a4paper,11pt]{article}
\usepackage[margin=1in]{geometry}
\usepackage{setspace,graphicx,url,hyperref}
\usepackage{amsmath,amssymb,amsfonts,amsthm,natbib}
\usepackage[linesnumbered,ruled]{algorithm2e}

\usepackage{amsmath,amssymb,amsfonts,amsthm,natbib}
\newcommand{\N}{\ensuremath{\mathbb{N}}}
\newcommand{\R}{\ensuremath{\mathbb{R}}}
\newcommand{\PP}{\ensuremath{\mathbb{P}}}
\newcommand{\E}{\ensuremath{\mathbb{E}}}
\newcommand{\I}{\ensuremath{\mathbb{I}}}
\newtheorem{lemma}{Lemma}

\begin{document}
\title{Optimal allocation of Monte Carlo simulations to multiple hypothesis tests}
\author{Georg Hahn}
\date{}
\maketitle

\begin{abstract}
Multiple hypothesis tests are often carried out in practice using p-value estimates obtained with bootstrap or permutation tests since the analytical p-values underlying all hypotheses are usually unknown. This article considers the allocation of a pre-specified total number of Monte Carlo simulations $K \in \N$ (i.e., permutations or draws from a bootstrap distribution) to a given number of $m \in \N$ hypotheses in order to approximate their p-values $p \in [0,1]^m$ in an optimal way, in the sense that the allocation minimises the total expected number of misclassified hypotheses. A misclassification occurs if a decision on a single hypothesis, obtained with an approximated p-value, differs from the one obtained if its p-value was known analytically. The contribution of this article is threefold: Under the assumption that $p$ is known and $K \in \R$, and using a normal approximation of the Binomial distribution, the optimal real-valued allocation of $K$ simulations to $m$ hypotheses is derived when correcting for multiplicity with the Bonferroni correction, both when computing the p-value estimates with or without a pseudo-count. Computational subtleties arising in the former case will be discussed. Second, with the help of an algorithm based on simulated annealing, empirical evidence is given that the optimal integer allocation is likely of the same form as the optimal real-valued allocation, and that both seem to coincide asympotically. Third, an empirical study on simulated and real data demonstrates that a recently proposed sampling algorithm based on Thompson sampling asympotically mimics the optimal (real-valued) allocation when the p-values are unknown and thus estimated at runtime.
\end{abstract}

\vspace{9pt}

\section{Introduction}
\label{sec:introduction}
Scientific studies are often evaluated by correcting for multiple comparisons using, for instance, the Bonferroni correction \citep{Bonferroni1936} or the procedures of \cite{Sidak1967}, \cite{Holm1979}, \cite{Simes1986}, \cite{Hochberg1988}, or \cite{BenjaminiHochberg1995}.

Although testing procedures such as the Bonferroni correction require exact knowledge of the p-value underlying each statistical test, p-values are usually not available analytically in practice and thus have to be approximated using Monte Carlo methods, for instance, bootstrap or permutation tests \citep{GandyHahn2014, GandyHahn2016, GandyHahn2017, SilvaAssuncao2018}. In the context of such Monte Carlo tests, the (analytical) p-value refers to the one obtained by integrating over the theoretical bootstrap distribution (in case of bootstrap tests) or by exhaustively generating all permutations (in case of permutation tests). This scenario is common in scientific studies involving real data, see, for instance, \cite{Tang2017}, \cite{ChenChen2017}, \cite{Wei2016}, or \cite{Shen2014} for recent testing applications involving Monte Carlo approximated p-values and \cite{Mrkvicka2017} or \cite{Pesarin2016} for Monte Carlo extensions of multiple testing.

When testing $m \in \N$ given hypotheses $H_{01},\ldots,H_{0m}$ in practice, a researcher is faced with the task of allocating a number of Monte Carlo simulations $K \in \N$ (which in practice is always finite) according to some criterion of choice in order to approximate the p-values of the $m$ hypotheses. It is assumed throughout the article that $H_{01},\ldots,H_{0m}$ are tested using independent test statistics. For simplicity, the classical Bonferroni correction is considered in the remainder of this article to correct for multiplicity, though it will be discussed how the results of this article extend to other multiple testing procedures. Recent examples for scientific works utilising the Bonferroni correction include \cite{Gallagher2018}, \cite{Zhang2017}, or \cite{Mestres2017}.

This article considers the optimal allocation of $K$ Monte Carlo simulations to $m$ hypotheses, in the sense that the allocation minimises the total expected number of misclassified hypotheses, under the assumption of a testing scenario in which independent test statistics are available and multiplicity is corrected with the Bonferroni correction.

To this end, two approaches are explored. First, it is assumed that the number of simulations to be allocated to each hypothesis is real-valued as opposed to integer-valued, and the Binomial distribution arising in the expression for the expected number of misclassifications is replaced by a normal approximation. This simplifies the problem and allows for the computation of gradients, thus making it possible to solve for the optimal real-valued allocation using the Karush-Kuhn-Tucker (KKT) formalism \citep{Karush1939, KuhnTucker1951}. This is done in two cases, precisely when p-value estimates are computed both with or without a pseudo-count \citep{DavisonHinkley1997}. In the former case, an optimal solution does not always exist and further computational subtleties arise, which will be discussed. Second, the computation of the optimal integer-valued allocation is attempted with the help of a scheme based on the simulated annealing (SA) algorithm \citep{Kirkpatrick1983} which, under conditions, allows to find integer solutions which converge to an optimal solution \citep{Henderson2003}.

Several algorithms to compute significant and non-significant hypotheses via approximated p-values are available in the literature. For instance, the method of \cite{BesagClifford1991}, the approaches of \cite{GuoPedadda2008} and \cite{Wieringen2008}, the \texttt{MCFDR} algorithm of \cite{Sandve2011}, the method of \cite{JiangSalzman2012} or the \texttt{MMCTest} algorithm of \cite{GandyHahn2014}. However, to the best of our knowledge, it is unclear how the allocation of Monte Carlo simulations computed by such algorithms available in the literature compares to the optimal allocation (in the above sense). Nevertheless, for the \texttt{QuickMMCTest} algorithm of \cite{GandyHahn2017}, a simulation study included in this article empirically demonstrates that its allocation asympotically mimics the optimal allocation of Monte Carlo simulations (as $K$ and $m$ go to infinity). This is of importance for practical applications: Generating simulations, for instance via permutations, can be computationally very expensive. An optimal (or nearly optimal) allocation of Monte Carlo simulations thus minimises computational resources while maximising the accuracy of the multiple testing result or makes the evaluation of real data possible in the first place.

For the special case of one hypothesis, a related field to the one of this article pertains to the sequential design of Monte Carlo testing while minimising the total number of simulations \citep{LanWittes1988, BesagClifford1991, Gandy2009, Fay2007, Silva2009, SilvaAssuncao2013}.

The article is organised as follows. Section~\ref{sec:optallocation} introduces the mathematical formulation of the problem under consideration (Section~\ref{sec:formulation}), simplifies it by allowing $K$ to be real-valued and by using a normal approximation of the Binomial distribution (Section~\ref{sec:normalapprox}), and solves for the optimal allocation using the KKT conditions (Section~\ref{sec:KKT}). Computational issues arising when solving the KKT conditions without (Section~\ref{sec:without_pseudo}) and with a pseudo-count (Section~\ref{sec:with_pseudo}) are discussed. As in practice any allocation is discrete, the related discrete optimisation problem is heuristically solved with the simulated annealing algorithm (Section~\ref{sec:SA}). A simulation study (Section~\ref{sec:simulation_study}) visualises how the optimal allocation of Monte Carlo simulations to $m$ hypotheses relates to their underlying p-value distribution (Section~\ref{sec:pvalues_and_opt}), and empirically demonstrates that the real-valued KKT as well as the integer-valued SA solutions are qualitatively similar both for a finite $K$ and asympotically (Section~\ref{sec:real_and_integer_allocation}). Moreover, in contrast to the aforementioned algorithms published in the literature, it is shown empirically that the \texttt{QuickMMCTest} algorithm \citep{GandyHahn2017} asympotically mimics the optimal (real-valued) allocation on simulated (Section~\ref{sec:quickmmctest}) and real data (Section~\ref{sec:pekowska}). The article concludes with a discussion in Section~\ref{sec:discussion}. Supplementary material containing $R$ code \citep{RDevelopmentCoreTeam2011RLa} to reproduce all figures is provided.

Throughout the article, let $\Vert \cdot \Vert$ denote the Euclidean norm of a vector. Let $\N_0 = \{0,1,2,\ldots\}$ denote the natural numbers including zero and let $\R_+ = \{ x \in \R: x>0 \}$ be the strictly positive part of the real line.

\section{The optimal allocation for the Bonferroni correction}
\label{sec:optallocation}
This section states a mathematical formulation of the optimisation problem of minimising the expected number of misclassified hypotheses (Section~\ref{sec:formulation}) and derives a real-valued solution using the KKT formalism (Section~\ref{sec:normalapprox}). Solving the KKT conditions (Section~\ref{sec:KKT}) is not straightforward if the p-value estimates for all hypotheses are computed with a pseudo-count \citep{DavisonHinkley1997}, and computational subtleties are discussed in Sections~\ref{sec:without_pseudo} and \ref{sec:with_pseudo}. Extensions of all results to other multiple testing procedures (apart from the Bonferroni correction) are discussed in Section~\ref{sec:other_procedures}.

\subsection{Formulation of the problem}
\label{sec:formulation}
A researcher is faced with testing $m \in \N$ hypotheses $H_{01},\ldots,H_{0m}$ for statistical significance using some test statistics and the Bonferroni correction at a given testing threshold $\alpha \in (0,1)$. Throughout the article, it is assumed that the test statistics for testing $H_{01},\ldots,H_{0m}$ are independent. Typically, $\alpha=t/m$, where $t$ is an uncorrected threshold such as $t=0.1$. Let $I:=\{ 1,\ldots,m \}$ and denote the unknown p-value underlying each hypothesis $H_{0i}$ as $p_i$, $i \in I$. The Bonferroni correction returning the indices of all rejected hypotheses is defined as $b(p,\alpha) = \{ i: p_i \leq \alpha \}$, where $p=(p_1,\ldots,p_m)$.

As the p-values are unknown, it is assumed that Monte Carlo methods are used to approximate them as $\hat{p}_i = (S_i+c)/(k_i+c)$, where $k_i$ is the total number of Monte Carlo simulations generated for $H_{0i}$ and $S_i$ is the number of exceedances over the observed test statistic (computed with some given data for each hypothesis) among those $k_i$ simulations, $i \in I$. The parameter $c \in \{0,1\}$ determines if a pseudo-count \citep{DavisonHinkley1997} is used in the numerator and denominator of $\hat{p}_i$ which bounds the estimates away from zero. Such a pseudo-count is recommended and commonplace in practice \citep{PhipsonSmyth2010}. Instead of generating Monte Carlo simulations, the number of significances can equivalently be modeled as $S_i \sim \text{Binomial}(k_i,p_i)$.

The hypothesis $H_{0i}$ is rejected if and only if $\hat{p}_i \leq \alpha$, where $i \in I$. All remaining hypotheses are non-rejected. The aim of this article is to find the optimal allocation of Monte Carlo simulations $k^\ast=(k_1^\ast,\ldots,k_m^\ast) \in \N_0^m$ to the hypotheses $H_{01},\ldots,H_{0m}$ which minimises the expected number of misclassifications, defined below, subject to the constraint that $\sum_{i=1}^m k_i^\ast=K$ for a given total number of simulations $K \in \N$ specified in advance.

Let $M_i = \{ \hat{p}_i > \alpha \wedge p_i \leq \alpha \} \cup \{ \hat{p}_i \leq \alpha \wedge p_i > \alpha \}$ be the event that hypothesis $H_{0i}$ is misclassified, that is the event that the unknown p-value $p_i$ of $H_{0i}$ and its estimate $\hat{p}_i$ lie on two different sides of the testing threshold. Using the event $M_i$, the total number of misclassifications can be expressed as $M = \sum_{i=1}^m \I_{M_i}$, where $\I$ is the indicator function. When allocating $k_i$ simulations to hypothesis $H_{0i}$ to estimate its p-value, the probability of a misclassification of hypothesis $H_{0i}$ is
\begin{align}
g_i(k_i) &:= \PP(M_i \mid k_i)
\nonumber
= \PP \left( \left. \frac{S_i+c}{k_i+c} > \alpha \right\vert p_i \right) \cdot \I(p_i \leq \alpha) +
\PP \left( \left. \frac{S_i+c}{k_i+c} \leq \alpha \right\vert p_i \right) \cdot \I(p_i > \alpha)\\
&= \PP \left( S_i > \alpha(k_i+c)-c \mid p_i \right) \cdot \I(p_i \leq \alpha) +
\PP \left( S_i \leq \alpha(k_i+c)-c \mid p_i \right) \cdot \I(p_i > \alpha),
\label{eq:g_i}
\end{align}
where the dependence of $g_i$ on $p_i$, $\alpha$ and $c$ is omitted for notational simplicity. The probability in \eqref{eq:g_i} is also called the \textit{resampling risk}, a popular error measure which many algorithms published in the literature on Monte Carlo hypothesis testing aim to control \citep{DavidsonMacKinnon2000, FayFollmann2002, Fay2007, Gandy2009, Kim2010, Ding2018}. The total expected number of misclassifications which occur when allocating $k=(k_1,\ldots,k_m)$ simulations to $H_{01},\ldots,H_{0m}$ is thus
\begin{align}
g(k) := \E(M|k) = \sum_{i=1}^m \PP(M_i|k_i) = \sum_{i=1}^m g_i(k_i),
\label{eq:g}
\end{align}
where the expectation is taken over the random $S_i \sim \text{Binomial}(k_i,p_i)$, $i \in I$. The goal is to minimise the total number of misclassifications $g(k)$ for a suitable choice $k^\ast \in \N_0^m$ with $\sum_{i=1}^m k_i^\ast=K$. The functions $g_i(k_i)$ go to zero as $k_i \rightarrow \infty$. This is to be expected as by the law of large numbers, each estimate $\hat{p}_i$ converges to the p-value $p_i$ as more Monte Carlo simulations are generated. To summarise, the constrained optimisation problem under investigation can be formalised as
\begin{align}
\min_{k \in \N_0^m} g(k) \qquad \text{subject to}~\sum_{i=1}^m k_i=K.
\label{eq:optprob}
\end{align}

\subsection{The optimal allocation for a normal approximation}
\label{sec:normalapprox}
The constrained optimisation in \eqref{eq:optprob} can be solved with the help of the KKT formalism. As derivatives are needed for KKT, \eqref{eq:g_i} is relaxed by allowing $k \in \R_+^m$ and by approximating $S_i \sim \text{Binomial}(k_i,p_i)$ in \eqref{eq:g_i} with a normal distribution with mean $k_i p_i$ and variance $k_i p_i (1-p_i)$. This yields
\begin{align*}
h_i(k_i) := \left[ 1-\Phi \left( \frac{k_i(\alpha-p_i)+c(\alpha-1)}{\sqrt{k_i p_i (1-p_i)}} \right) \right] \cdot \I(p_i \leq \alpha)
+ \Phi \left( \frac{k_i(\alpha-p_i)+c(\alpha-1)}{\sqrt{k_i p_i (1-p_i)}} \right) \cdot \I(p_i > \alpha),
\end{align*}
where $\Phi$ is the cumulative distribution function of the standard normal distribution and where it was used that $\PP(X \leq x) = \Phi((x-\mu)/\sigma)$ for $X \sim N(\mu,\sigma^2)$.

Now, $g_i(k_i) \approx h_i(k_i)$ for all $i \in I$ and consequently, \eqref{eq:g} can be approximated as $g(k) \approx h(k):=\sum_{i=1}^m h_i(k_i)$. By the de Moivre--Laplace theorem, the ratio of $g_i(k_i)$ to $h_i(k_i)$ tends to one as $k_i \rightarrow \infty$ for each $i \in I$. Thus as $K \rightarrow \infty$, given each hypothesis receives an amount of $k_i$ of those $K$ simulations with $k_i \rightarrow \infty$, the approximation $h$ will be very accurate. For small $K$, however, $h$ might be a poor approximation of $g$ and \eqref{eq:optprob} should be solved directly for an integer solution as attempted in Section~\ref{sec:SA} (this is easier for small $K$ than for large ones).

The derivative of each $h_i$ (which is a function of $k_i$ only), $i \in I$, is given by
\begin{align}
\frac{\partial h_i}{\partial k_i}
\nonumber
=& - \frac{k_i(\alpha-p_i)-c(\alpha-1)}{2 k_i \sqrt{k_i p_i (1-p_i)}}
\cdot \phi \left( \frac{k_i(\alpha-p_i)+c(\alpha-1)}{\sqrt{k_i p_i (1-p_i)}} \right) \cdot \I(p_i \leq \alpha)\\
&+ \frac{k_i(\alpha-p_i)-c(\alpha-1)}{2 k_i \sqrt{k_i p_i (1-p_i)}}
\cdot \phi \left( \frac{k_i(\alpha-p_i)+c(\alpha-1)}{\sqrt{k_i p_i (1-p_i)}} \right) \cdot \I(p_i > \alpha),
\label{eq:derivh}
\end{align}
where $\phi$ is the density function of the standard normal distribution. The partial derivative $\partial h/\partial k_i$ depends on $k_i$ only, thus allowing to essentially separate the optimisation problem into $m$ problems, each finding an optimal number of $k_i^\ast$ simulations for hypothesis $H_{0i}$.

The function $h$ needs to be optimised under the constraints $k_i > 0$ for $i \in I$ and $\sum_{i=1}^m k_i = K$, meaning that each hypothesis receives a positive number of simulations and that the total number of simulations allocated equals $K$. The optimal solution $k^\ast$ minimising $h$ satisfies the Lagrangian associated with the constrained optimisation problem, given by
\begin{align}
\nabla h(k^\ast) = \sum_{i=1}^m \nu_i \nabla u_i(k^\ast) + \lambda^\ast \nabla v(k^\ast),
\label{eq:kkt1}
\end{align}
where $u_i(k)=-k_i$ and $v(k) = K-\sum_{i=1}^m k_i$, $k=(k_1,\ldots,k_m)$. The functions $u_i$ encode the constraint $k_i > 0$ (primal feasibility) with $\nu_i \geq 0$ (dual feasibility) and satisfy $\nu_i u_i(k^\ast)=0$ (complementary slackness), where $i \in I$.

Complementary slackness and the condition $k_i > 0$ imply that $\nu_i=0$ for all $i \in I$. As each partial derivative $\partial h / \partial k_i$ only depends on $k_i$, \eqref{eq:kkt1} simplifies to
\begin{align}
\frac{\partial h}{\partial k_i}(k_i^\ast) = -\lambda^\ast
\label{eq:kkt2}
\end{align}
for $i \in I$ and an optimal value $\lambda^\ast > 0$ ensuring $\sum_{i=1}^m k_i^\ast = K$.

\subsection{Computational considerations when solving the KKT conditions}
\label{sec:KKT}
Finding the optimal value $\lambda^\ast$ in \eqref{eq:kkt2} and the corresponding $k^\ast$ which satisfies $\sum_{i=1}^m k_i^\ast = K$ is straightforward if no pseudo-count is used when computing p-value estimates ($c=0$ in \eqref{eq:derivh}) and more challenging with a pseudo-count ($c=1$ in \eqref{eq:derivh}). In particular, the optimal solution allocating $K$ Monte Carlo simulations to $m$ hypotheses might not always exist in the latter case. The following two Sections~\ref{sec:without_pseudo} and \ref{sec:with_pseudo} provide computational details.

\subsubsection{P-value estimates without a pseudo-count}
\label{sec:without_pseudo}
\begin{figure}[t]
\centering
\includegraphics[width=0.5\textwidth]{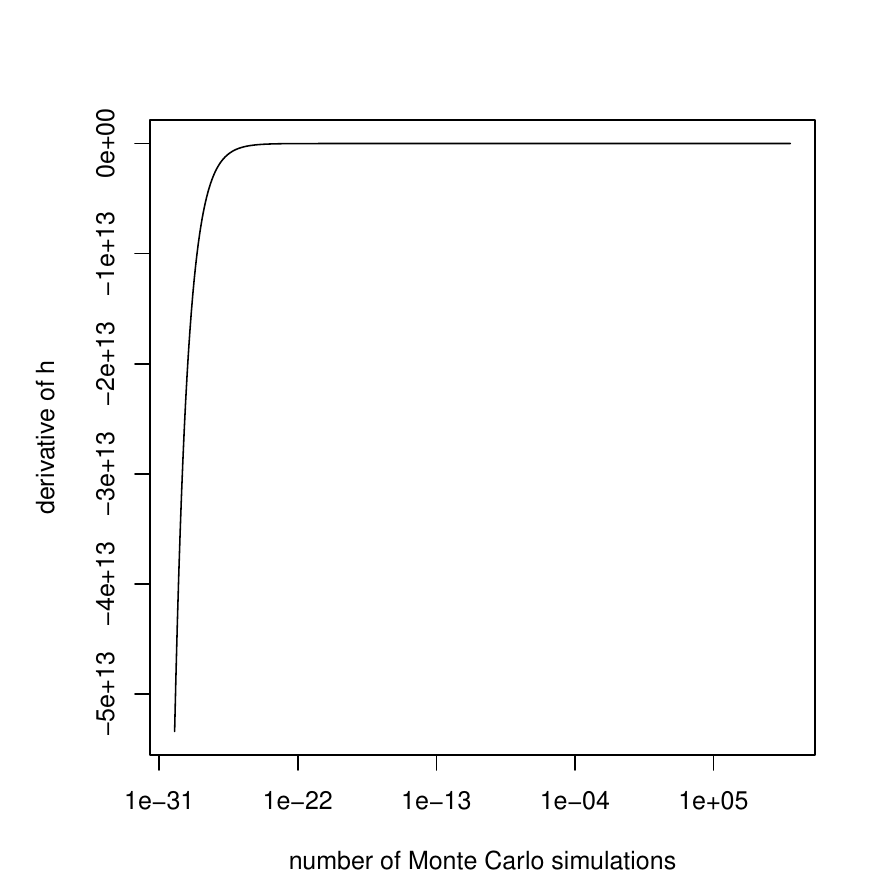}
\caption{Qualitative behaviour of $\partial h / \partial k_i$ as a function of $k_i$ for p-value estimates without a pseudo-count ($c=0$ in \eqref{eq:derivh}). The behaviour is identical for p-values both below and above the testing threshold $\alpha$.\label{fig:plotderiv_pc0}}
\end{figure}
If no pseudo-count is used ($c=0$ in \eqref{eq:derivh}), the qualitative behaviour of $\partial h / \partial k_i$ is as depicted in Figure~\ref{fig:plotderiv_pc0}. The derivative is negative (given $p_i \neq \alpha$) and strictly increases to zero as $k_i \rightarrow \infty$. This is proven in Lemma~\ref{lemma:strictly} in Section~\ref{sec:proof}.

Solving for the optimal allocation is thus straightforward and computationally efficient: First, suppose a value $\lambda > 0$ and the index $i \in I$ of a particular hypothesis are given and the task is to find the value of $k_i$ such that $(\partial h / \partial k_i) (k_i) = -\lambda$, formalised as the following problem:
\begin{align}
P_{i,\lambda}: ~\text{return}~k_i \in \R_+~\text{satisfying}~(\partial h / \partial k_i) (k_i) = -\lambda.
\label{eq:getk}
\end{align}
To solve \eqref{eq:getk} a double binary search can be employed: Starting with an arbitrary starting value (e.g., $k_i^0 = 1$), double or halve $k_i^0$ until an interval $[k_i^L,k_i^R]$ is obtained such that $-\lambda \in [(\partial h / \partial k_i)(k_i^L),(\partial h / \partial k_i)(k_i^R)]$. A binary search applied to that interval then finds the solution $k_i$ of $P_{i,\lambda}$ in logarithmic time (in the size of the search space). Both steps rely on the fact that $\partial h / \partial k_i$ are strictly increasing for all $i \in I$. Since the size (length) of the search space is bounded by $K$, this operation takes $O(\log K)$ for each $i \in I$.

Since the derivatives $\partial h / \partial k_i$ are negative but strictly increasing for all $i \in I$, lower (higher) values of $\lambda>0$ yield larger (smaller) $k_i$ across all $i \in I$. Thus in a second step, choose an arbitrary starting value $\lambda_0$ (e.g., $\lambda_0 = 1$), determine the corresponding vector $k=(k_1,\ldots,k_m)$ by solving $P_{i,\lambda_0}$ for all $i \in I$ and check if $\sum_{i=1}^m k_i - K$ is positive or negative. If it is positive, $\lambda_0$ will be doubled (thus decreasing $-\lambda_0$) and otherwise halved until an interval $[\lambda^L,\lambda^R]$ is found which contains the optimal value $\lambda^\ast$ whose corresponding solution $k^\ast=(k_1^\ast,\ldots,k_m^\ast)$ of $P_{i,\lambda^\ast}$ for all $i \in I$ satisfies $\sum_{i=1}^m k_i^\ast = K$. A binary search then finds $\lambda^\ast$ within $[\lambda^L,\lambda^R]$ in logarithmic time. The total effort for finding both $\lambda^\ast$ and the optimal allocation of Monte Carlo simulations can thus be expressed as $O(m \log(\lambda^R-\lambda^L) \log K)$. Section~\ref{sec:simulation_study} shows that for \textit{typical} testing scenarios, $\lambda^\ast$ is of the order of around $10^{-10}$ to $10^{-5}$. In principle, any number of $K \geq 0$ simulations can be allocated to the $m$ hypotheses in this way.

\subsubsection{P-value estimates with a pseudo-count}
\label{sec:with_pseudo}
\begin{figure}[t]
\centering
\begin{minipage}{0.49\textwidth}
\includegraphics[width=\textwidth]{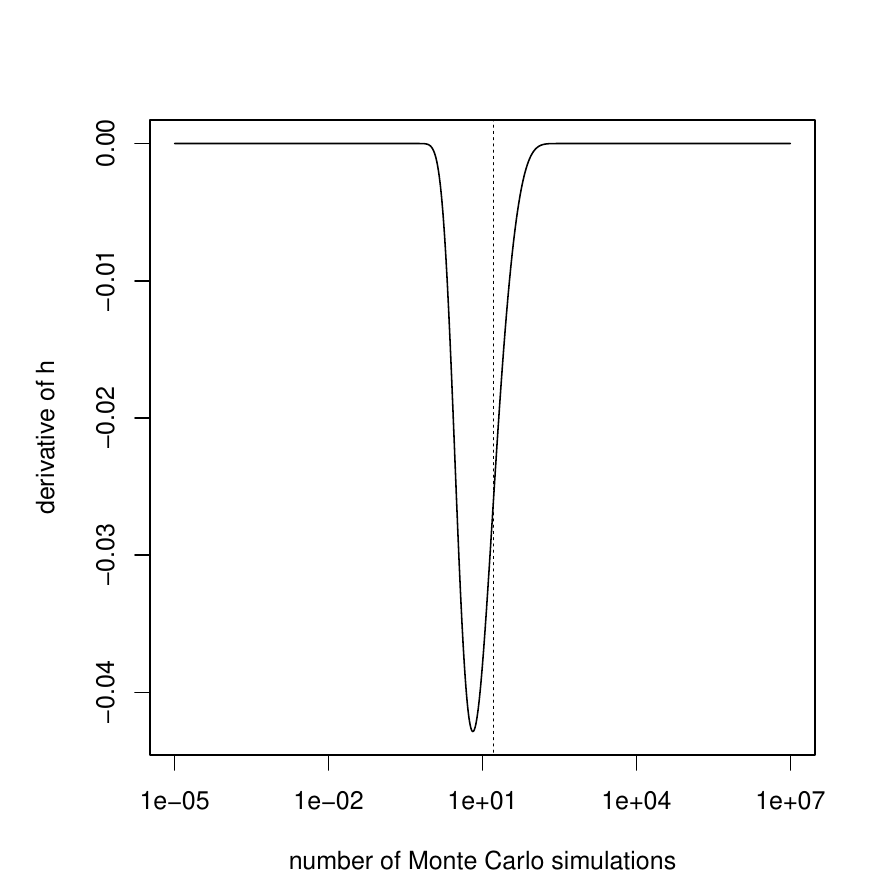}
\end{minipage}~
\begin{minipage}{0.49\textwidth}
\includegraphics[width=\textwidth]{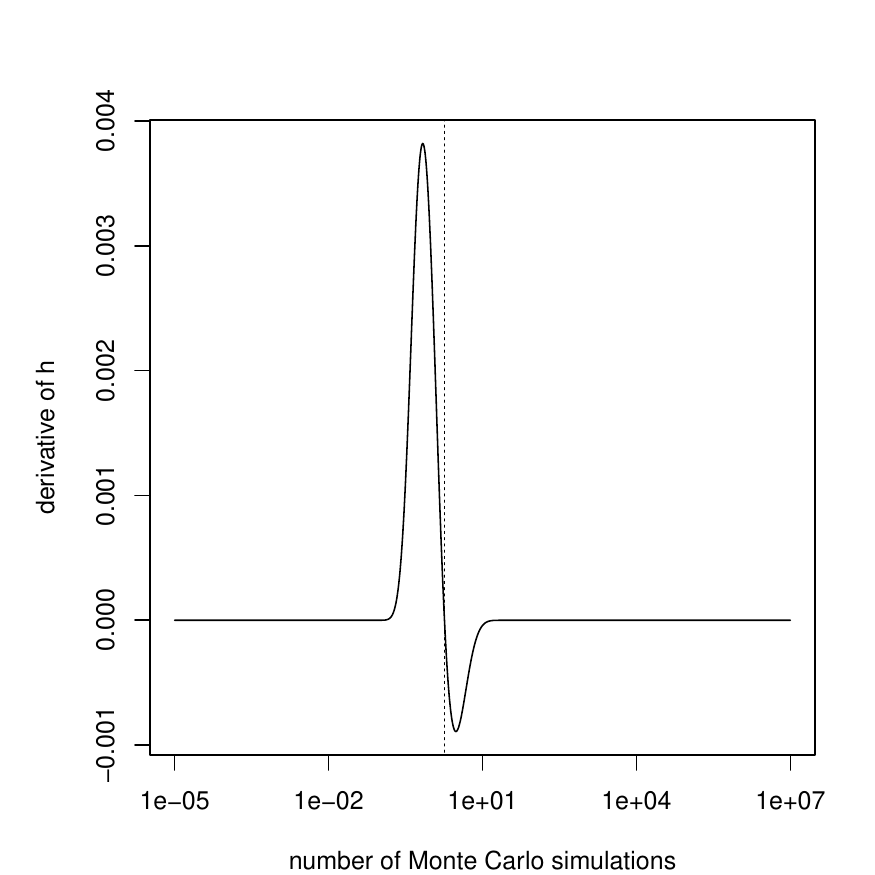}
\end{minipage}
\caption{Qualitative behaviour of $\partial h / \partial k_i$ as a function of $k_i$ for p-value estimates with a pseudo-count ($c=1$ in \eqref{eq:derivh}) and for a p-value below (left) and above (right) the testing threshold $\alpha$. The guess $\gamma_i$ defined in \eqref{eq:guess} is indicated with a dotted vertical line.\label{fig:plotderiv_pc1}}
\end{figure}
When using a pseudo-count, finding $k_i$, $i \in I$, for a given $\lambda$ as well as finding the optimal $\lambda^\ast$ becomes more challenging. This is due to the fact that the derivatives $\partial h / \partial k_i$ are not strictly increasing any more and neither do they attain all values in $(-\infty,0)$. Figure~\ref{fig:plotderiv_pc1} shows examples of the qualitative behaviour of $\partial h / \partial k_i$ for a p-value below (left) and above (right) the testing threshold $\alpha$.

Since in both cases, the derivative is still negative in a limited region, it might nevertheless be possible to find a $\lambda^\ast>0$ that satisfies the KKT conditions. To this end, first determine the range of admissible values of $k_i$ for each hypothesis $H_{0i}$, $i \in I$, defined as the range of values for which $\partial h / \partial k_i$ is negative and strictly increasing. To apply binary search as done for solving \eqref{eq:getk}, find the range extending from the unique minimum $\mu_i$ (see Figure~\ref{fig:plotderiv_pc1}) of the derivative $\partial h / \partial k_i$ to the maximal number of Monte Carlo simulations $K$ to be allocated, a natural upper bound of $k_i$.

However, since $K$ can be large it is non-trivial to locate $\mu_i$ (or even any non-zero value of the derivative subject to machine precision), and it is helpful to have a \textit{guess} of the location of the minimum. This guess can be computed as the point at which the derivative \eqref{eq:derivh} changes from being positive to negative. In \eqref{eq:derivh}, the function $\phi$ is always strictly positive, so it suffices to consider the zero of its prefactor, which is easily computed as
\begin{align}
\frac{k_i(\alpha-p_i)-c(\alpha-1)}{2 k_i \sqrt{k_i p_i (1-p_i)}} = 0 \qquad \Rightarrow \qquad \gamma_i := k_i = \frac{c(\alpha-1)}{\alpha-p_i}.
\label{eq:guess}
\end{align}
The quantity $\gamma_i$ is referred to as the guess for the minimum of hypothesis $H_{0i}$ (depicted as dotted vertical line in Figure~\ref{fig:plotderiv_pc1}). Note that \eqref{eq:guess} remains valid for the case $c=0$ in which the minimum is at $k_i=0$ for all $i \in I$ (see Figure~\ref{fig:plotderiv_pc0}).

Using $\gamma_i$, a binary search determines the two points to the left and to the right of $\gamma_i$ at which the derivative is first non-zero (subject to machine precision). This gives a search window for $\mu_i$ (computable in $O(\log K)$ time), and $\mu_i$ is then efficiently found with, for instance, a golden section minimisation procedure \citep{Kiefer1953,AvrielWilde1966} as implemented in the $R$ function \texttt{optimise} \citep{RDevelopmentCoreTeam2011RLa}. The complexity of the golden section search is $O(\log(1/\epsilon))$, where $\epsilon$ is the computational precision \citep{Luenberger2003}. Once $\mu_i$ is found, the resulting range of admissible values for $k_i$ can be set as $[k_i^L,k_i^R] := [\mu_i,K]$ for each $i \in I$.

After $[k_i^L,k_i^R]$ is determined for each hypothesis $H_{0i}$, $i \in I$, the optimal value $\lambda^\ast$ has to be found. For this, a search window for the binary search on $\lambda$ is again needed: this search window precisely consists of all those values $\lambda$ for which $P_{i,\lambda}$ has a solution within the admissible range $[k_i^L,k_i^R]$ \textit{for all} indices $i \in I$. However, finding such an interval for $\lambda^\ast$ is not always possible as shown in the following paragraphs and empirically in Section~\ref{sec:real_and_integer_allocation}.

Since by construction, $\partial h / \partial k_i$ is strictly increasing within $[k_i^L,k_i^R]$ for all $i \in I$, the window $[k_i^L,k_i^R]$ of admissible values for each $k_i$ can be translated to a search window for the admissible values of $\lambda$ that correspond to it. To be precise, this search interval is $[\lambda_i^L,\lambda_i^R] := [-(\partial h / \partial k_i)(k_i^R), -(\partial h / \partial k_i)(k_i^L)]$ for all $i \in I$, since large (small) values of $k_i$ correspond to low (high) values of $\lambda>0$.

It remains to compute the intersection of all $[\lambda_i^L,\lambda_i^R]$, $i \in I$, that is the range $[\lambda^L,\lambda^R] := [ \max_i \lambda_i^L, \min_i \lambda_i^R ]$ of $\lambda$ values which guarantees that $P_{i,\lambda}$ has a solution for all $i \in I$. However, it can happen that a few single hypotheses cause $\lambda^L > \lambda^R$, precisely those in $J := \{j \in I: \lambda_j^L > \lambda^R \vee \lambda_j^R < \lambda^L \}$. In this case, a global optimal allocation might not exist. However, the existence of an optimal allocation can be guaranteed again if the hypotheses in $J$ are removed from $I$.

If $[\lambda^L,\lambda^R]$ is an interval of non-zero length, then by construction $P_{i,\lambda}$ can be solved for all $i \in I$ and for any $\lambda \in [\lambda^L,\lambda^R]$. In particular, for any given $\lambda \in [\lambda^L,\lambda^R]$, $P_{i,\lambda}$ is solved with a binary search on the individual search window $[k_i^L,k_i^R]$ for each $i \in I$ as in Section~\ref{sec:without_pseudo}, and the optimal $\lambda^\ast$ is likewise found with a binary search within $[\lambda^L,\lambda^R]$, again with total computational effort $O(m \log(\lambda^R-\lambda^L) \log K)$.

Conversely, the final range $[\lambda^L,\lambda^R]$ of admissible $\lambda$ values can be translated into the minimal and maximal number of Monte Carlo simulations that can be allocated in an optimal way for the given p-values $p$. The minimal and maximal numbers are $\sum_{i=1}^m k_i^L$ and $\sum_{i=1}^m k_i^R$, respectively, where $k_i^L$ is the solution of $P_{i,\lambda^R}$ for all $i \in I$ and $k_i^R$ is the solution of $P_{i,\lambda^L}$ for all $i \in I$ (the minimal optimal allocation is obtained for the largest admissible value of $\lambda>0$ and likewise for the maximal optimal allocation). These bounds on the minimal and maximal numbers of simulations which can be allocated in an optimal way will be used in Section~\ref{sec:real_and_integer_allocation}.

\subsection{Extension to other multiple testing procedures}
\label{sec:other_procedures}
The previous derivations specifically addressed the Bonferroni correction with constant threshold $\alpha$. However, since the computation of the optimal allocation assumes full knowledge of both the p-values $p$ and the threshold $\alpha$, this is not a restriction. This is due to the fact that the result of any multiple testing procedure $P(p,\alpha)$ can equivalently be obtained by applying the Bonferroni correction to $p$ with constant threshold set to the p-value of the last rejected hypothesis by $P$.

\section{A simulated annealing algorithm to attempt the computation of the optimal integer allocation}
\label{sec:SA}
Section~\ref{sec:optallocation} considered the computation of the optimal allocation of Monte Carlo simulations under the assumption that $k \in \R_+^m$ using a normal approximation of the Binomial distribution in \eqref{eq:g_i}. In this section, a scheme based on simulated annealing (SA) of \cite{Kirkpatrick1983} is derived in order to attempt to solve \eqref{eq:optprob} directly. To this end, Section~\ref{sec:behaviour_gi} first considers the behaviour of the functions $g_i$ occurring in \eqref{eq:optprob}. Section~\ref{sec:proposalSA} discusses the design of a suitable proposal function for SA and Section~\ref{sec:algoSA} introduces the actual SA algorithm. Although the optimality of the found integer allocation cannot be guaranteed any more, a simulation study (Section~\ref{sec:real_and_integer_allocation}) will later give empirical evidence that the optimal integer allocation is likely of the same form as the optimal real-valued KKT allocation.

\subsection{Qualitative behaviour of the probability of misclassifications}
\label{sec:behaviour_gi}
\begin{figure}[t]
\centering
\begin{minipage}{0.49\textwidth}
\includegraphics[width=\textwidth]{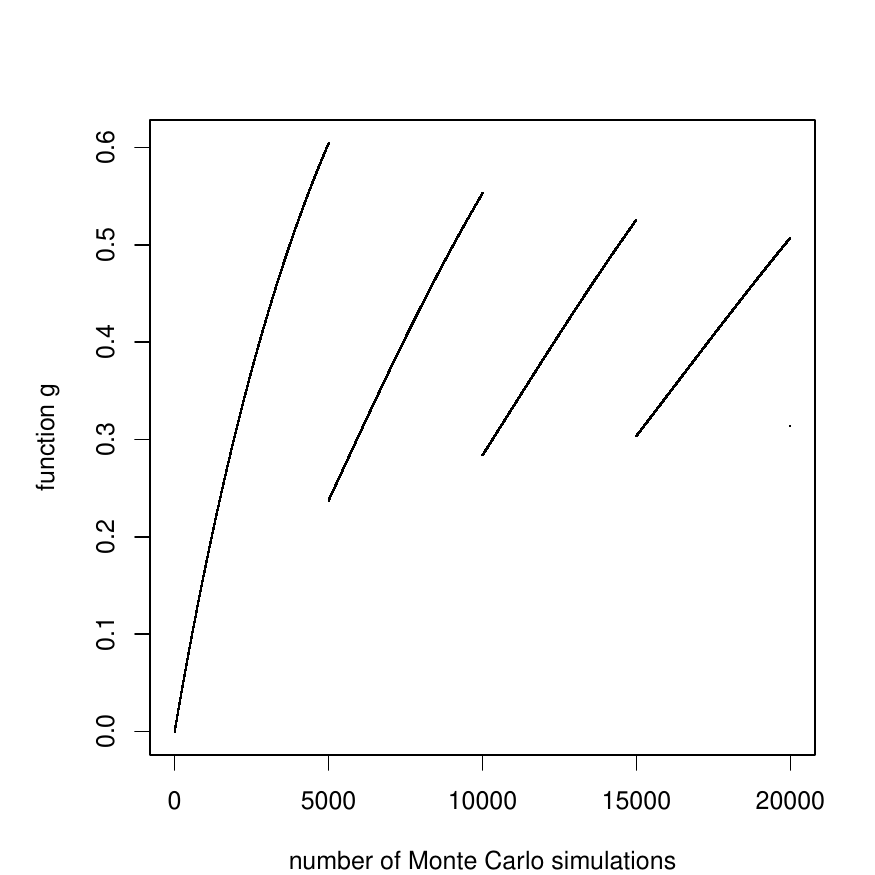}
\end{minipage}~
\begin{minipage}{0.49\textwidth}
\includegraphics[width=\textwidth]{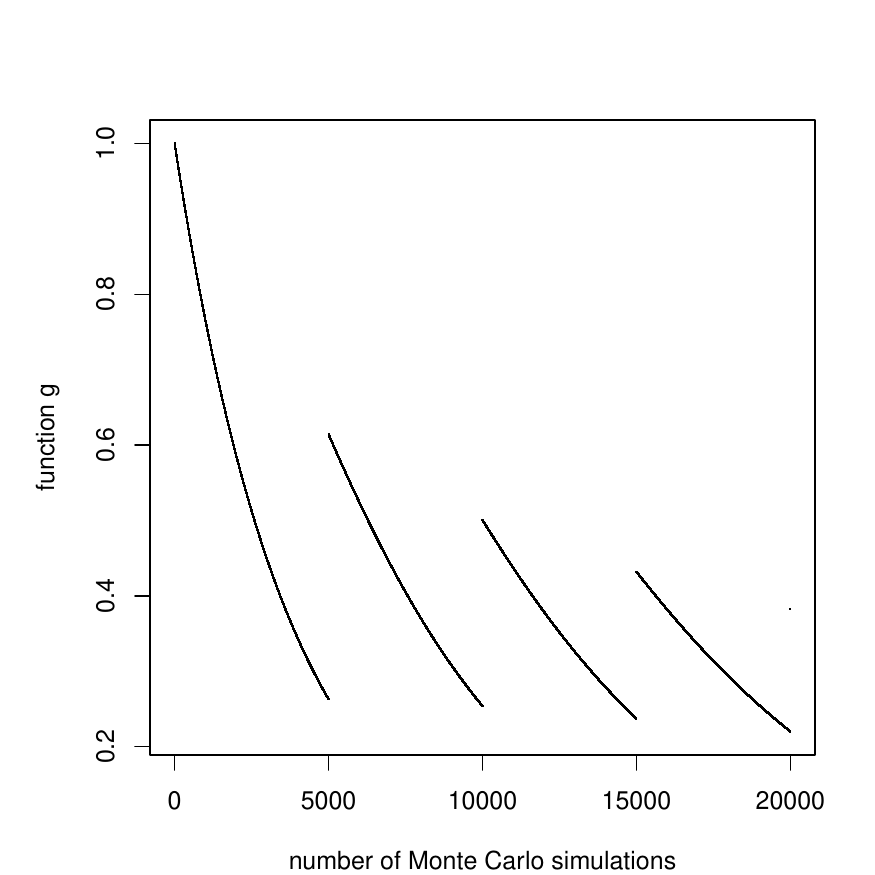}
\end{minipage}
\caption{Qualitative behaviour of the expected number of misclassifications $g_i$ (see \eqref{eq:g_i}) as a function of $k_i$ when computing p-value estimates without a pseudo-count ($c=0$ in \eqref{eq:derivh}) and for a p-value below (left) and above (right) the testing threshold. Each new ``branch'' begins at a multiple of $1/\alpha$ plus one simulation, that is at $\nu/\alpha+1$ for $\nu \in \N$ (here, $\alpha=1/5000$).\label{fig:exactg_pc0}}
\end{figure}
In order to motivate the design of the SA proposal in Section~\ref{sec:proposalSA}, consider the behaviour of the function $g_i$ depicted exemplarily in Figure~\ref{fig:exactg_pc0} for a p-value $p_i$ below (left) and above (right) the testing threshold $\alpha=1/5000$.

In Figure~\ref{fig:exactg_pc0}, the absence of a pseudo-count ($c=0$ in \eqref{eq:derivh}) is crucial. A rejection is obtained if a p-value estimate is below $\alpha$, where generating $0$ simulations is defined to yield a p-value of $0$ and hence a correct rejection. Thus when generating less than $5000$ simulations for a p-value below $\alpha$ (left), only the case of $0$ simulations leads to a sure rejection (and thus to a correct decision, that is a probability of misclassification of zero). Generating more than $0$ simulations results in a non-zero probability of observing at least $1$ exceedance (in which case the hypothesis under consideration will be erroneously non-rejected), thus increasing the probability of a misclassification. On reaching $1/\alpha=5000$ simulations, both $0$ or $1$ exceedances will lead to a rejection and hence to a correct decision, thus causing the expected number of misclassifications to drop again.

For a p-value above $\alpha$ (right), the inverse effect happens. Generating no simulations leads to a p-value of $0$ (resulting in a rejection) and thus to a sure misclassification. Generating more simulations decreases the expected number of misclassifications. On reaching $1/\alpha=5000$ simulations, observing both $0$ and $1$ exceedances leads to a rejection and thus to a misclassification, causing the expected number of misclassifications to increase again.

\begin{figure}[t]
\centering
\begin{minipage}{0.49\textwidth}
\includegraphics[width=\textwidth]{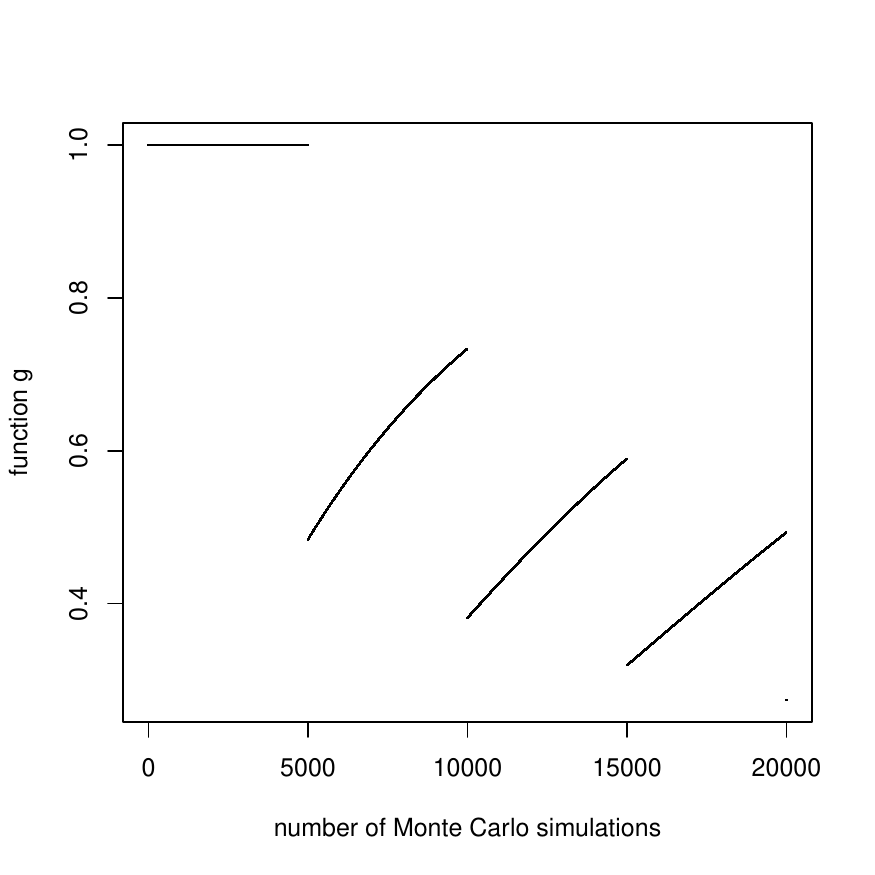}
\end{minipage}~
\begin{minipage}{0.49\textwidth}
\includegraphics[width=\textwidth]{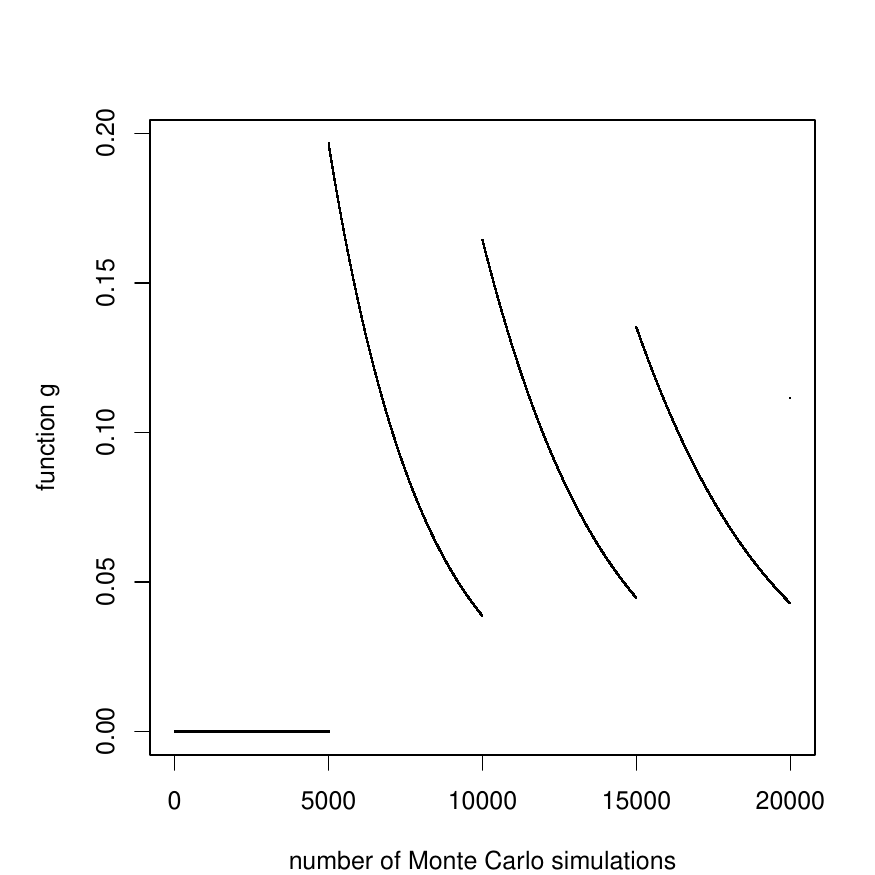}
\end{minipage}
\caption{Qualitative behaviour of the expected number of misclassifications $g_i$ (see \eqref{eq:g_i}) as a function of $k_i$ when computing p-value estimates with a pseudo-count ($c=1$ in \eqref{eq:derivh}) and for a p-value below (left) and above (right) the testing threshold. Each new ``branch'' begins at a multiple of $1/\alpha$ simulations (here, $\alpha=1/5000$).\label{fig:exactg_pc1}}
\end{figure}

Figure~\ref{fig:exactg_pc1} depicts a similar behaviour of the number of misclassifications when computing p-value estimates with a pseudo-count ($c=1$ in \eqref{eq:derivh}). Here, generating no simulations results in a p-value estimate of $1$ and thus in a sure non-rejection, and likewise generating any number of simulations less than $1/\alpha-1$ leads to a p-value estimate strictly above $\alpha$ irrespective of the number of observed exceedances. As expected, in Figures~\ref{fig:exactg_pc0} and \ref{fig:exactg_pc1} the probability of a misclassification vanishes for any hypothesis $H_{0i}$ as $k_i \rightarrow \infty$, $i \in I$.

\subsection{The choice of the simulated annealing proposal}
\label{sec:proposalSA}
\begin{algorithm}[t]
\caption{\texttt{proposal}}
\label{algo:proposal}
  \SetKwInOut{Input}{input}
  \SetKwFor{Loop}{repeat}{}{end}
  \Input{$k$, $S$, $\alpha$, $j_0 \leftarrow \lfloor 1/\alpha \rfloor$\;}
  Set $j \leftarrow j_0$ or $j \leftarrow 1$ with probability $0.5$ each\;
  \If{$\exists i \in S$: $k_i \geq j$}{
    Uniformly draw $u$ from $\{ i \in S: k_i \geq j \}$\;
    $k_u \leftarrow k_u - j$\;
    $v \leftarrow \arg\max_{i \in I} g_i(k_i)-g_i(k_i+j)$\;
    $k_v \leftarrow k_v + j$\;
  }
  \Return{$k$}\;
\end{algorithm}
When no pseudo-count is used ($c=0$ in \eqref{eq:derivh}), according to Figure~\ref{fig:exactg_pc0}, it is only sensible to allocate batches of $1/\alpha$ Monte Carlo simulations to hypotheses with p-values below the testing threshold. The same applies to hypotheses with p-values above the threshold, except when a hypothesis receives less than $1/\alpha$ simulations, in which case the probability of a misclassification for that hypothesis decreases as more simulations are being allocated to it (corresponding to the first branch in Figure~\ref{fig:exactg_pc0}, right). A similar observation holds true for p-values computed with a pseudo-count (Figure~\ref{fig:exactg_pc1}).

Based on this observation, Algorithm~\ref{algo:proposal} offers a sensible proposal for an SA algorithm. The function \texttt{proposal} takes as inputs the current allocation of Monte Carlo simulations $k=(k_1,\ldots,k_m)$ to all $m$ hypotheses, a finite set $S \subseteq I$ indicating which hypotheses ought to be eligible for a fine tuning of their currently allocated number of simulations, the threshold $\alpha$ and a jump size $j_0$ which is preset to $\lfloor 1/\alpha \rfloor$.

Algorithm~\ref{algo:proposal} works as follows. First, the actual jump size $j$ to be used is set to either $j_0$ or $1$ with probability $0.5$ each. This ensures that the SA algorithm will not be restricted to allocating batches of $j_0$ simulations only, but will be able to fine-tune the allocation. If there exists at least one hypothesis from which $j$ simulations can be taken for re-allocation, then an index $u$ is drawn uniformly among all those hypotheses currently receiving at least $j$ simulations, $j$ is substracted from $k_u$, and the $j$ simulations are allocated in a greedy fashion to the hypothesis which yields the largest decrease in its expected number of misclassifications. The greedy allocation is not necessary but employed here to speed up the slow SA scheme. Alternatively, the $j$ simulations can also be allocated to any randomly chosen hypothesis.

\subsection{A simulated annealing approach}
\label{sec:algoSA}
\begin{algorithm}[t]
\caption{\texttt{SA}}
\label{algo:SA}
  \SetKwInOut{Input}{input}
  \SetKwFor{Loop}{repeat}{}{end}
  \Input{$k_0$, $F$, $P$, $s^{\max}$, $\beta \leftarrow 10^{-4}$\;}
  $k \leftarrow k_0$\;
  \For{$s \leftarrow 1$ \KwTo $s^{\max}$}{
    $T \leftarrow \beta/\log(s+1)$\;
    \label{eq:logtemp}
    $\tilde{k} \leftarrow P(k)$\;
    $a \leftarrow \min\left( 1, \exp( (F(k)-F(\tilde{k}))/T ) \right)$\;
    Uniformly draw $u$ from $[0,1]$\;
    \lIf{$u < a$}{$k \leftarrow \tilde{k}$}
    \label{eq:acceptance}
  }
  \Return{$k$}\;
\end{algorithm}
Using the proposal in Algorithm~\ref{algo:proposal}, the minimisation in \eqref{eq:optprob} can now be attempted with the help of a simple SA scheme \citep{Kirkpatrick1983} given in Algorithm~\ref{algo:SA}.

Algorithm~\ref{algo:SA} works as follows. In each iteration, the proposal function $P$ is called with the current allocation $k$ (initialised with some allocation $k_0$) and its proposed new allocation is saved in $\tilde{k}$. Both the current allocation $k$ and the proposed new allocation $\tilde{k}$ are evaluated on $F$ (the function to be minimised, in our case $F=g$ defined in \eqref{eq:g}) and a standard SA acceptance probability $a$ is computed. If $F(\tilde{k}) \leq F(k)$ then the argument of the exponential function will be non-negative and thus $a=1$, leading to a sure acceptance of the proposal $\tilde{k}$ in line~\ref{eq:acceptance}. If $F(\tilde{k})>F(k)$ then the proposal $\tilde{k}$ will only be accepted with probability $a$ in line~\ref{eq:acceptance}. Since in this case the proposal $\tilde{k}$ actually increases the objective function, SA has the ability to leave local minima. The aforementioned steps are repeated over a pre-specified number of iterations $s^{\max}$. The last accepted proposal $k$ is returned as the output of Algorithm~\ref{algo:SA}.

To turn SA into a steepest descent optimiser over time, the argument of the exponential function is weighted by $1/T$ using a \textit{temperature} $T$ which is decreased in every step (line~\ref{eq:logtemp}). Employing a logarithmic decrease ensures, under conditions, that SA will find the global minimum of $F$ as $s^{\max} \rightarrow \infty$ \citep{Henderson2003}.

The initial allocation $k^\text{init}$ can be chosen as follows. If $c=0$ then it is sensible to set $k_i^\text{init}$ to a multiple of $1/\alpha$ plus one for all indices in $\{i \in I: p_i \leq \alpha\}$, and likewise $k_i^\text{init}$ is set to a multiple of $1/\alpha$ for all $\{i \in I: p_i > \alpha\}$ (see Figure~\ref{fig:exactg_pc0}). Similarly, for $c=1$, $k_i^\text{init}$ is set to a multiple of $1/\alpha$ for all indices in $\{i \in I: p_i \leq \alpha\}$, and to a multiple of $1/\alpha$ minus one otherwise. If at the end, $\sum_{i=1}^m k_i^\text{init} < K$, one Monte Carlo simulation each is added to a randomly drawn entry among $\{i \in I: p_i > \alpha\}$ ($\{i \in I: p_i \leq \alpha\}$) for $c=0$ ($c=1$) until $\sum_{i=1}^m k_i^\text{init} = K$ is satisfied. Afterwards, the two calls
\begin{align*}
&k_1 := \texttt{SA}(k_0=k^\text{init}, F=g(k), P=\texttt{proposal}(k,S=\{i \in I: p_i > \alpha\},\alpha=\alpha), s^{\max}=10^6),\\
&k_2 := \texttt{SA}(k_0=k_1, F=g(k), P=\texttt{proposal}(k,S=\{i \in I: p_i \leq \alpha\},\alpha=\alpha), s^{\max}=10^6),
\end{align*}
optimise the allocation of Monte Carlo simulations both above and below the threshold, where $g$ is as defined in \eqref{eq:g}. Since $k^\text{init} \in \N_0^m$, $\sum_{i=1}^m k_i^\text{init} = K$, and since the proposal in Algorithm~\ref{algo:proposal} only performs swaps of an integer number of simulations between hypotheses, it is guaranteed that $k_1,k_2 \in \N_0^m$ and that both sum up to $K$. The integer allocation of Monte Carlo simulations in vector $k_2$ is returned as the SA approximation of the optimal allocation $k^\ast$ solving \eqref{eq:optprob}.

\section{Simulation study}
\label{sec:simulation_study}
This section discusses the optimal allocation for a given distribution of p-values (Section~\ref{sec:pvalues_and_opt}) and shows that the optimal real-valued and integer allocations are qualitatively similar (Section~\ref{sec:real_and_integer_allocation}), both for a finite $K$ and asympotically. Section~\ref{sec:quickmmctest} empirically demonstrates that while adaptively generating Monte Carlo simulations at runtime to approximate the unknown p-values of multiple hypotheses, the \texttt{QuickMMCTest} algorithm of \cite{GandyHahn2017} allocates simulations in a way that asympotically coincides with the optimal (real-valued) allocation. A comparison on a real dataset to the other algorithms listed in Section~\ref{sec:introduction} is given in Section~\ref{sec:pekowska}.

\subsection{Relationship between p-value distribution and optimal allocation}
\label{sec:pvalues_and_opt}
\begin{figure}[t]
\centering
\includegraphics[width=0.5\textwidth]{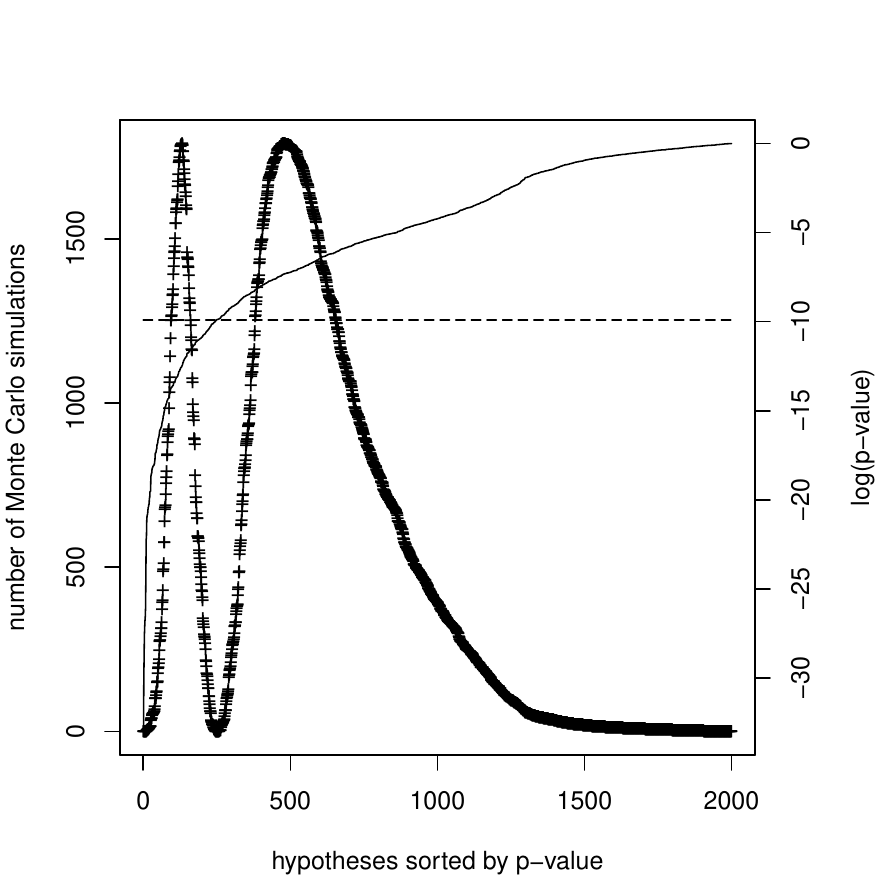}
\caption{Exemplary p-value distribution for $m=2000$ hypotheses (solid line, p-values on the logarithmic right axis), testing threshold $\alpha=0.1/m$ (horizontal dashed line) and number of simulations allocated to each hypothesis (crosses, numbers given on the left axis).\label{fig:pvalues_and_opt}}
\end{figure}

This section visualises how the distribution of p-values is related to the optimal (real-valued) allocation of Monte Carlo simulations to all hypotheses. For this, $m$ p-values are drawn from the mixture distribution proposed in \cite{Sandve2011}: The mixture distribution consists of a proportion $\pi_0$ of true null hypotheses drawn from a uniform distribution in $[0,1]$, and a proportion $1-\pi_0$ drawn from a $Beta(0.25,25)$ distribution. This distribution resembles p-value distributions observed in real data studies (e.g., genome testing) and has already been used in \cite{GandyHahn2014,GandyHahn2016,GandyHahn2017}.

Figure~\ref{fig:pvalues_and_opt} shows the distribution of $m=2000$ p-values drawn with $\pi_0=0.4$ (logarithmic axis on the right). The relatively low proportion $\pi_0$ of true null hypotheses is employed in this and the following sections to better visualise the p-value distribution and the resulting allocation of simulations. The testing threshold is set to $\alpha=0.1/m$. Using the known p-values, the optimal allocation of $K=10^6$ simulations was computed as described in Section~\ref{sec:without_pseudo} (without a pseudo-count, case $c=0$) and added to Figure~\ref{fig:pvalues_and_opt} as crosses (numbers given on the left axis). The optimal allocation was found with $\lambda^\ast=6.76 \cdot 10^{-5}$.

As expected, and as confirmed by empirical studies \citep{GandyHahn2017}, hypotheses with very small or very large p-values only require relatively few Monte Carlo simulations for stable decisions. As the p-value distribution approaches a neighbourhood of the testing threshold from either side, more simulations are required to minimise the total number of misclassifications in the optimal allocation. However, it turns out that in this example, hypotheses with p-values too close to the testing threshold are not worth being invested too many simulations in since the numbers required for a (reasonably) low probability of a misclassification are too large. Therefore, in the optimal allocation, those misclassifications are traded in for being able to use the unspent simulations on other hypotheses instead which are slightly further away from the threshold.

\begin{figure}[t]
\centering
\begin{minipage}{0.32\textwidth}
\includegraphics[width=\textwidth]{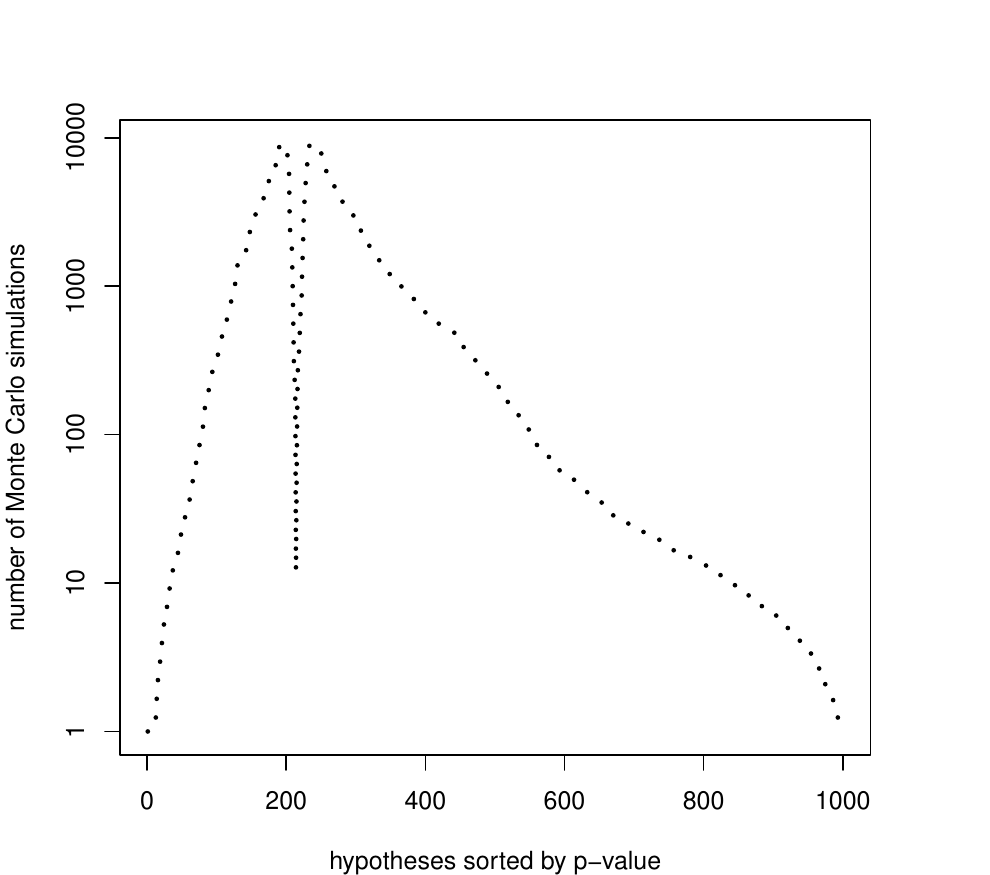}
\end{minipage}~
\begin{minipage}{0.32\textwidth}
\includegraphics[width=\textwidth]{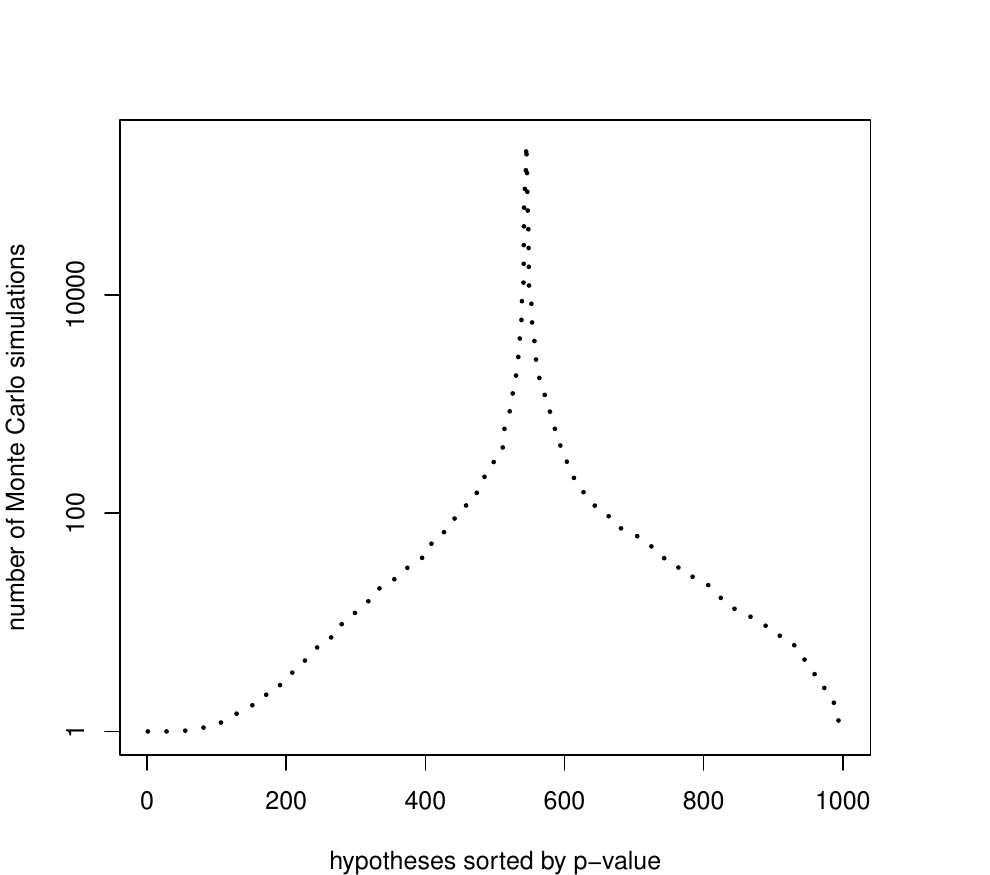}
\end{minipage}~
\begin{minipage}{0.32\textwidth}
\includegraphics[width=\textwidth]{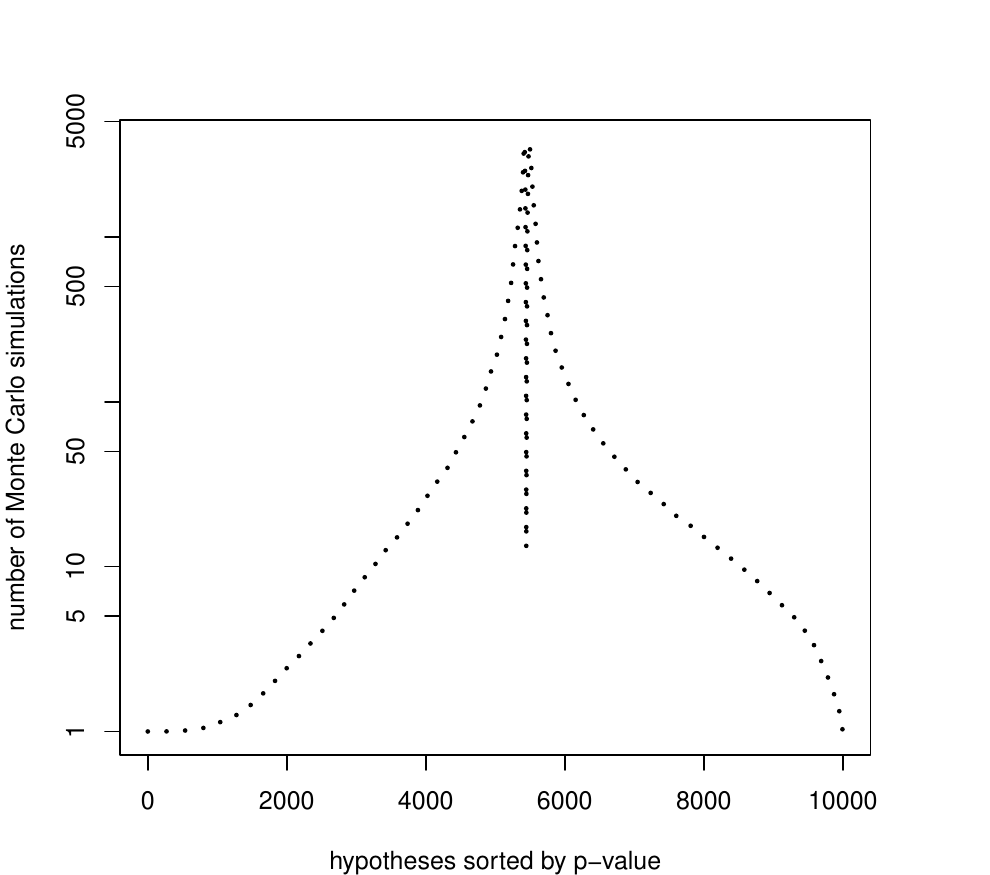}
\end{minipage}
\caption{Progression of optimal allocations $k^\ast$: $m=1000$ and $\alpha=10^{-3}$ with $\min_{i \in I} |p_i-\alpha| \propto 10^{-5}$ (left), $m=1000$ and $\alpha=10^{-1}$ with $\min_{i \in I} |p_i-\alpha| \propto 10^{-3}$ (middle), $m=10000$ and $\alpha=10^{-1}$ with $\min_{i \in I} |p_i-\alpha| \propto 10^{-5}$ (right). The y-axes display $\log(1+k^\ast)$.\label{fig:progression}}
\end{figure}

It is not always the case that hypotheses too close to the threshold receive less Monte Carlo simulations than those in an immediate neighbourhood. Figure~\ref{fig:progression} shows the optimal real-valued allocations for three scenarios using $K=10^6$ and sets of p-values drawn from the \cite{Sandve2011} distribution with parameter $\pi_0=0.5$. For $m=1000$ hypotheses and threshold $\alpha=10^{-3}$ (left), hypotheses close to the threshold do not receive many simulations. However, when increasing the threshold to $\alpha=10^{-1}$ (middle), this effect disappears and the optimal allocation invests most simulations in the hypotheses closest to the threshold. When increasing the number of hypotheses to $m=10000$ while keeping $\alpha=10^{-1}$ constant, hypotheses close to the threshold again receive less simulations.

This can be explained as follows. Suppose the optimal $\lambda^\ast$ was known. The optimal allocation consists of the $k_i^\ast$ for each hypothesis $H_{0i}$, $i \in I$, which satisfies \eqref{eq:kkt2}. The derivative $\partial h_i/\partial k_i$ in \eqref{eq:kkt2} is of the order
$$\frac{\partial h_i}{\partial k_i} \propto \frac{\alpha-p_i}{\sqrt{k_i p_i(1-p_i)}} \cdot \phi \left( \frac{\sqrt{k_i}(\alpha-p_i)}{\sqrt{p_i(1-p_i)}} \right)$$
for both cases $p_i \leq \alpha$ and $p_i>\alpha$, where it was used that $c=0$. The quantity of interest is $\alpha-p_i$: For a p-value $p_i>\alpha$ ($p_i<\alpha$) further away from the threshold, $\alpha-p_i$ is of the order of $p_i$ (of the order of $\alpha$), and $k_i^\ast$ has to be sufficiently large to ensure $(\partial h_i/\partial k_i)(k_i^\ast) = -\lambda^\ast$. If $p_i$ is very close to $\alpha$, $\alpha-p_i$ can be magnitudes smaller than both $p_i$ and $\alpha$, and the $k_i^\ast$ satisfying $(\partial h_i/\partial k_i)(k_i^\ast) = -\lambda^\ast$ need only be relatively small.

Figure~\ref{fig:progression} confirms this picture: The p-value distribution of \cite{Sandve2011} consists of a proportion $1-\pi_0$ of very small p-values from a $Beta(0.25,25)$ distribution, which cluster close to zero, and a proportion $\pi_0$ of p-values drawn from a uniform distribution which scatter within the entire interval $[0,1]$. In Figure~\ref{fig:progression} (left), $\alpha = 10^{-3}$ falls within the p-values from the Beta cluster, causing $\min_{i \in I} |p_i-\alpha|$ to be small (of the order of $10^{-5}$) and the optimal allocation to spend less Monte Carlo simulations right at the threshold. When increasing the threshold to $\alpha=10^{-1}$ (middle) while keeping $m$ fixed, $\alpha$ now falls within the uniform p-values which are scattered over the entire interval $[0,1]$, thus causing $\min_{i \in I} |p_i-\alpha| \propto 10^{-3}$ to increase to the point that hypotheses at the threshold receive most simulations. However, $\min_{i \in I} |p_i-\alpha|$ can be made smaller again simply by increasing $m$, for instance to $m=10000$ (thus increasing the p-value density in $[0,1]$ which leads to $\min_{i \in I} |p_i-\alpha| \propto 10^{-5}$), causing the optimal allocation to again spend less simulations at the threshold (right).

\subsection{Comparison of the optimal real-valued and integer allocations}
\label{sec:real_and_integer_allocation}
\begin{figure}[t]
\centering
\begin{minipage}{0.49\textwidth}
\includegraphics[width=\textwidth]{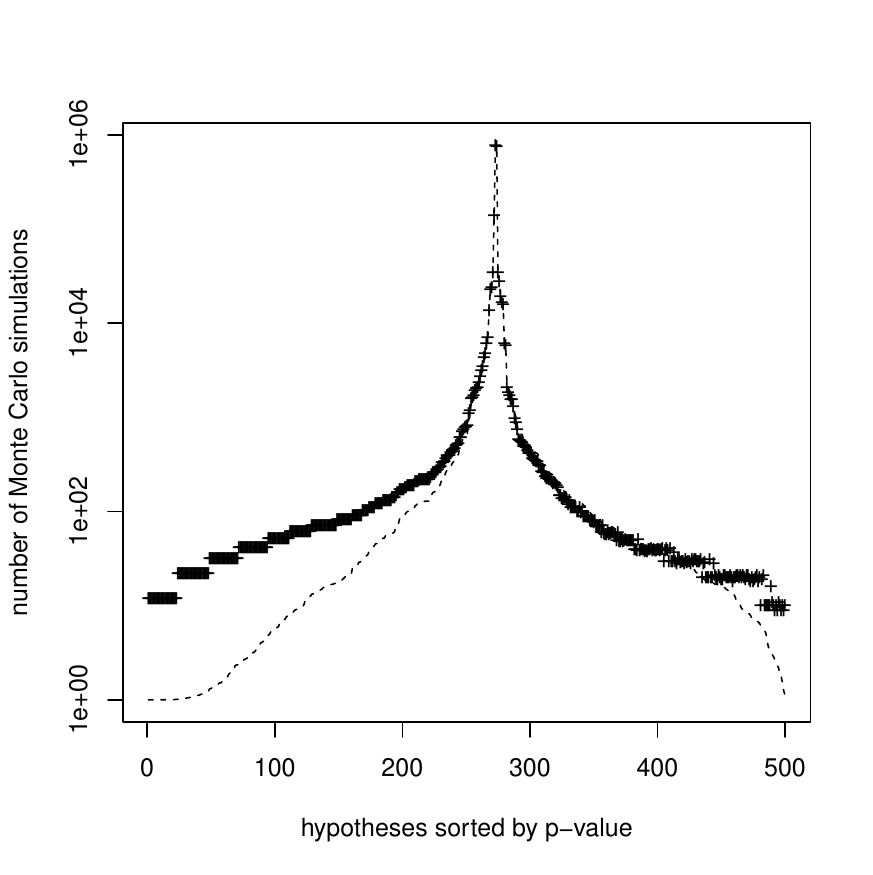}
\end{minipage}~
\begin{minipage}{0.49\textwidth}
\includegraphics[width=\textwidth]{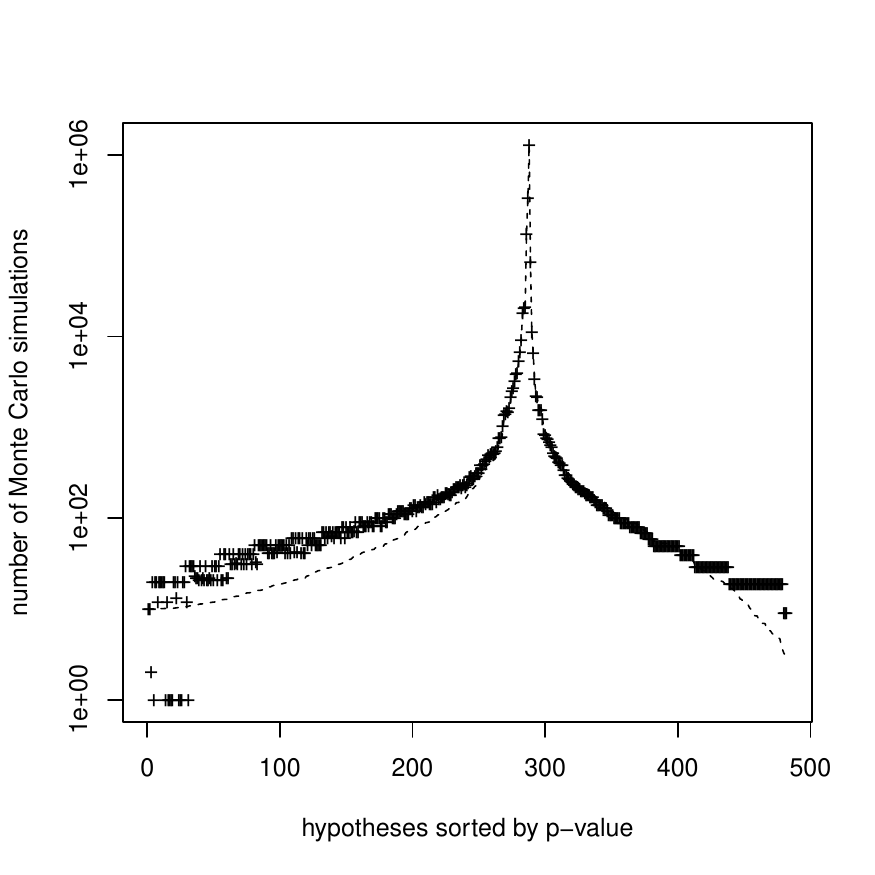}
\end{minipage}
\caption{Optimal KKT allocation $k^\ast$ computed as in Sections~\ref{sec:without_pseudo} and \ref{sec:with_pseudo} (dashed line) and integer solution $k_\text{SA}$ computed as in Section~\ref{sec:SA} (crosses), both without (left) and with a pseudo-count (right). The y-axes display $\log(1+k^\ast)$ and $\log(1+k_\text{SA})$, respectively.\label{fig:KKT_and_SANN}}
\end{figure}

Figure~\ref{fig:KKT_and_SANN} compares the optimal real-valued allocation $k^\ast$ (dashed line), computed both without (left) and with (right) a pseudo-count (Sections~\ref{sec:without_pseudo} and \ref{sec:with_pseudo}), to the SA integer solution $k_\text{SA}$ of Section~\ref{sec:SA} (crosses). The p-value distribution is again the mixture distribution of \cite{Sandve2011} (see Section~\ref{sec:pvalues_and_opt}) with parameters $m=500$, $\pi_0=0.5$, $K=2 \cdot 10^6$ and $\alpha=0.1$.

As seen in Figure~\ref{fig:KKT_and_SANN}, the two allocations are qualitatively similar. The KKT solution (dashed line) is smoother (since it is allowed to allocate a real-valued number of Monte Carlo simulations to each hypothesis) and has a more pronounced spike at the threshold, whereas SA seems to allocate less simulations to hypotheses at the threshold and more simulations to those hypotheses further away from the threshold.

For the case $c=0$ (Figure~\ref{fig:KKT_and_SANN}, left), the optimal allocation of simulations $k^\ast$ is found for $\lambda^\ast=1.92 \cdot 10^{-11}$ (in this example run). For the case $c=1$ (Figure~\ref{fig:KKT_and_SANN}, right), the computation of the optimal allocation was initially not possible since the hypotheses with ranks $483,\ldots,500$ (based on the ordered $m=500$ p-values) caused the search interval $[\lambda^L,\lambda^R]$ for the optimal $\lambda^\ast$ to be empty (see Section~\ref{sec:with_pseudo}). Removing the hypotheses with indices in $J:=\{483,\ldots,500\}$ led to $[\lambda^L,\lambda^R]=[1.35 \cdot 10^{-11}, 9.40 \cdot 10^{-10}]$ which translates to the range of simulations $[1930003,3007833]$ that can be allocated in an optimal way using KKT. Since $K=2 \cdot 10^6$ falls within that range, the optimal $\lambda^\ast=7.11 \cdot 10^{-10}$ was efficiently found.

\begin{figure}[t]
\centering
\includegraphics[width=0.5\textwidth]{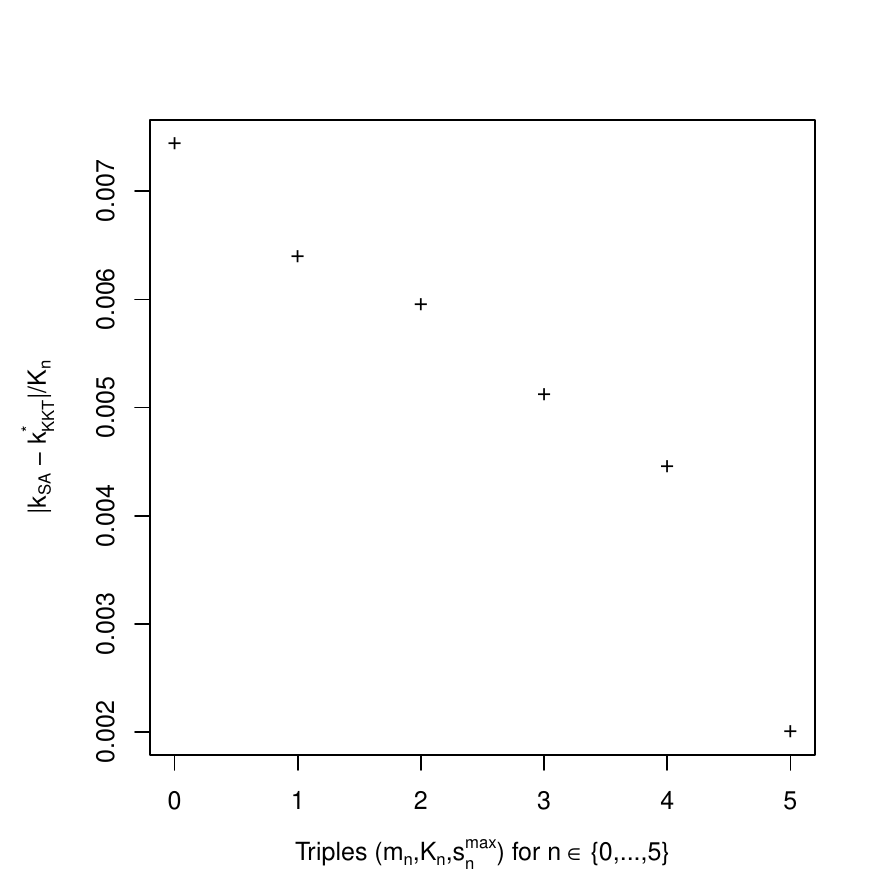}
\caption{Relative difference in vectors $k_\text{SA}$ (allocation of SA of Section~\ref{sec:SA}) and $k_\text{KKT}^\ast$ (optimal KKT allocation computed as in Section~\ref{sec:without_pseudo}) as a function of $(m_n,K_n,s^{\max}_n)$, $n \in \{0,\ldots,5\}$.\label{fig:KKT_and_SANN_asymptotics}}
\end{figure}

To quantify the similarity between the SA and KKT solutions, Figure~\ref{fig:KKT_and_SANN_asymptotics} compares both allocations as the number of hypotheses $m$, the number of Monte Carlo simulations $K$, and the number of iterations $s^{\max}$ of SA increase. Naturally, when increasing $m$, it is necessary to increase $K$ as well to ensure that enough simulations are available for all $m$ hypotheses, and likewise with increasing parameters $m$ and $K$ the SA algorithm requires more iterations $s^{\max}$ to compute allocations. The parameters $m$, $K$ and $s^{\max}$ are thus increased together as $(m_n,K_n,s^{\max}_n) = (50 \cdot 2^n, 10^5 \cdot 2^n, 10^6 \cdot 2^n)$ for $n \in \{0,\ldots,5\}$. As SA allocates an integer number of simulations, it seems unreasonable to assume that the SA allocation in vector $k_\text{SA}$ and the optimal real-valued KKT allocation $k_\text{KKT}^\ast$ will coincide in an $L_2$ sense. Instead, Figure~\ref{fig:KKT_and_SANN_asymptotics} shows $\Vert k_\text{SA}-k_\text{KKT}^\ast \Vert/K_n$, the relative difference in $L_2$ norm between the two allocation vectors which is normalised with respect to the number of simulations $K_n$ spent. Each datapoint is the median of $100$ repetitions. Figure~\ref{fig:KKT_and_SANN_asymptotics} indicates that the normalised $L_2$ difference between both allocation vectors seems to decrease.

\subsection{Comparison to Thompson sampling in the QuickMMCTest algorithm}
\label{sec:quickmmctest}
This section compares the optimal KKT allocation computed in Section~\ref{sec:KKT} with the allocation returned by the \texttt{QuickMMCTest} algorithm of \cite{GandyHahn2017}.

\texttt{QuickMMCTest} can be used to compute a decision (rejection or non-rejection) for a given set of hypotheses with unknown p-values based solely on Monte Carlo simulations. The algorithm aims to use more simulations for hypotheses with an (analytical and unknown) p-value close to the testing threshold (thus having a less stable decision, in the sense that their decision switches from being rejected to non-rejected if the data were analysed repeatedly), and less simulations for hypotheses with a p-value further away from the threshold (thus having a more stable decision). To compute a stability measure, \texttt{QuickMMCTest} starts with a uniform prior on the p-value of each hypothesis and updates a Beta-Binomial model for each p-value as more simulations are generated. Based on Thompson sampling \citep{Thompson1933,AgrawalGoyal2012}, in each iteration, a new p-value is drawn from each posterior and new rejections and non-rejections are computed for the sampled p-value distribution. Repeating this step several times gives a measure of how stable the current decision on each hypothesis is, which is then used to compute weights employed to allocate a new batch of simulations. Further details can be found in \cite{GandyHahn2017}.

\begin{figure}[t]
\centering
\begin{minipage}{0.49\textwidth}
\includegraphics[width=\textwidth]{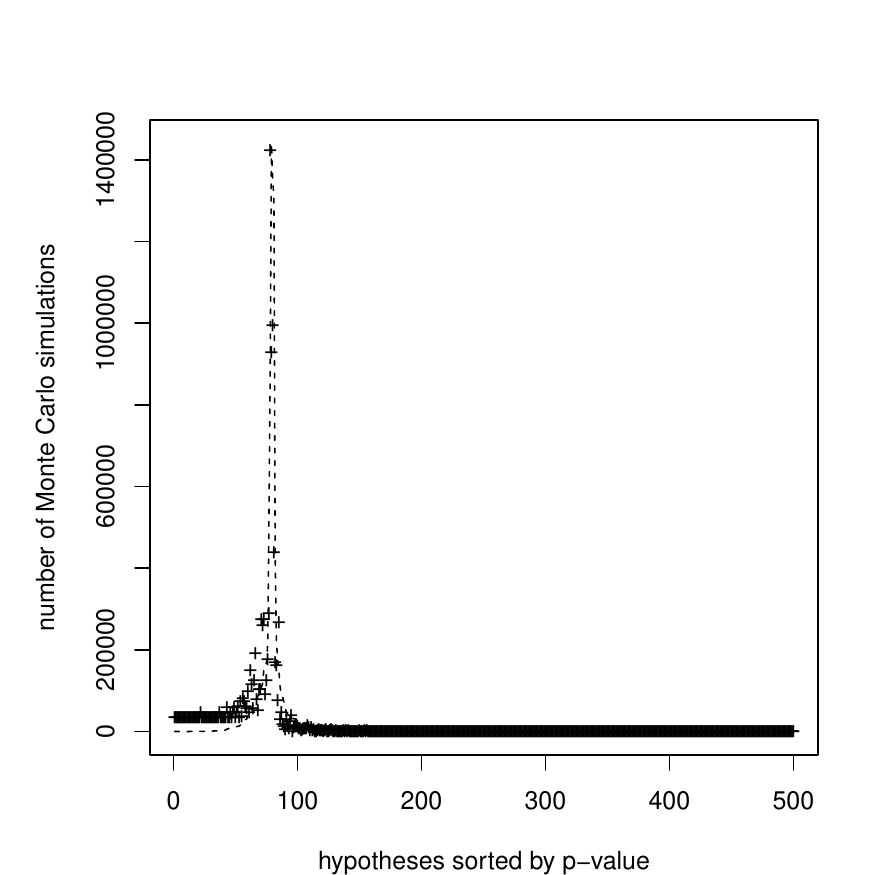}
\end{minipage}~
\begin{minipage}{0.49\textwidth}
\includegraphics[width=\textwidth]{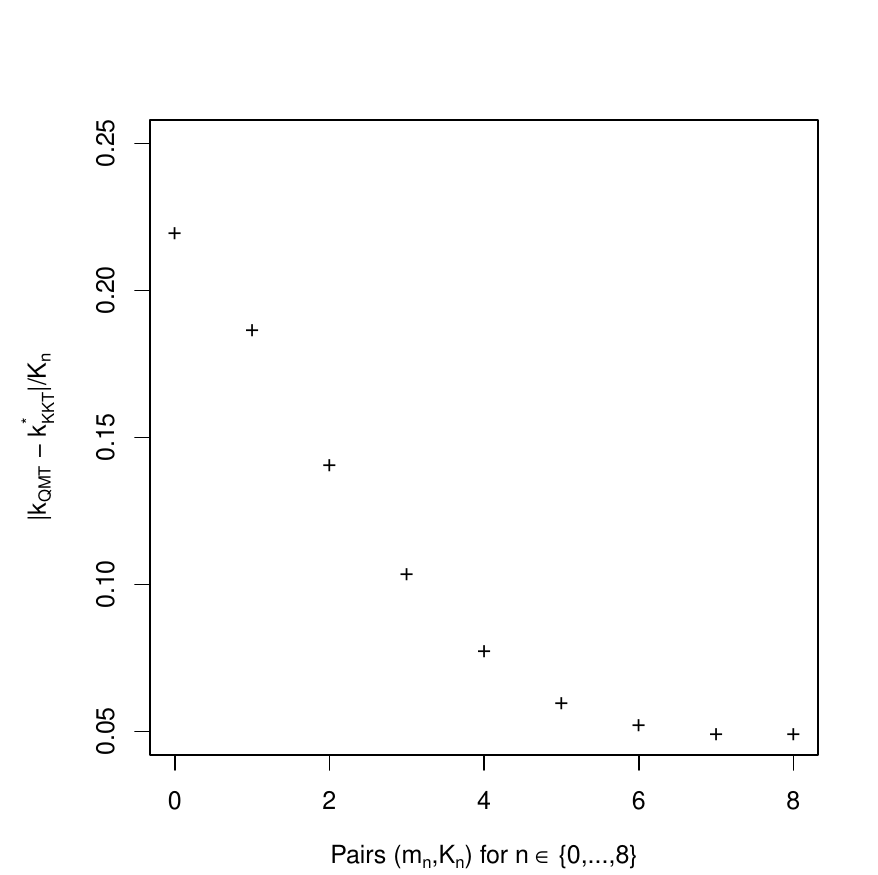}
\end{minipage}
\caption{Left: Optimal KKT allocation computed as in Section~\ref{sec:without_pseudo} (dashed line) and integer allocation of \texttt{QuickMMCTest} (crosses). No pseudo-count ($c=0$ in \eqref{eq:derivh}). Right: Relative difference in vectors $k_\text{QMT}$ (allocation of \texttt{QuickMMCTest}) and $k_\text{KKT}^\ast$ (optimal KKT allocation computed as in Section~\ref{sec:without_pseudo}) as a function of $(m_n,K_n)$, $n \in \{0,\ldots,8\}$.\label{fig:QMT_and_KKT}}
\end{figure}

Figure~\ref{fig:QMT_and_KKT} (left) shows the number of Monte Carlo simulations allocated to each hypothesis in an example run of \texttt{QuickMMCTest} (crosses), as well as the optimal KKT allocation (dashed line). For this, $m=500$ p-values were generated using the \cite{Sandve2011} distribution (see Section~\ref{sec:pvalues_and_opt}) with $\pi_0=0.5$ and the two allocations were computed with $K=10^7$ and $\alpha=0.1/m$. To ensure a fine-tuned allocation, \texttt{QuickMMCTest} was run with parameter $\Delta=10m$ (that is with a low average number of $10$ simulations spent per hypothesis in each iteration), all other parameters were kept at the default values given in \cite{GandyHahn2017}. As visible in Figure~\ref{fig:QMT_and_KKT} (left), \texttt{QuickMMCTest} manages to allocate simulations without knowledge of the p-values in a qualitatively similar fashion to the optimal allocation.

To quantify this similarity, as in Section~\ref{sec:real_and_integer_allocation}, both the \texttt{QuickMMCTest} allocation $k_\text{QMT}$ and the optimal KKT allocation $k_\text{KKT}^\ast$ are compared as both the number of hypotheses $m$ and the number of simulations $K$ increase. Like SA, \texttt{QuickMMCTest} is a probabilistic method which allocates integer numbers of simulations and thus Figure~\ref{fig:QMT_and_KKT} (right) shows $\Vert k_\text{QMT}-k_\text{KKT}^\ast \Vert/K_n$, the relative difference in $L_2$ norm between the two allocation vectors which is normalised with respect to the number of simulations $K_n$ spent. The parameters $m$ and $K$ are increased together as $(m_n,K_n) = (50 \cdot 2^n, 10^5 \cdot 2^n)$ for $n \in \{0,\ldots,8\}$. Each datapoint is the median of $100$ repetitions. As visible in the plot, the normalised $L_2$ difference between both allocation vectors seems to decrease.

A repetition of this experiment for a higher proportion of null hypotheses $\pi_0=0.9$ can be found in Section~\ref{sec:highpi}, again confirming the asympotic similarity of the two allocation vectors.

\subsection{Comparison to other algorithms on a real dataset}
\label{sec:pekowska}
\begin{figure}[t]
\centering
\begin{minipage}{0.32\textwidth}
\includegraphics[width=\textwidth]{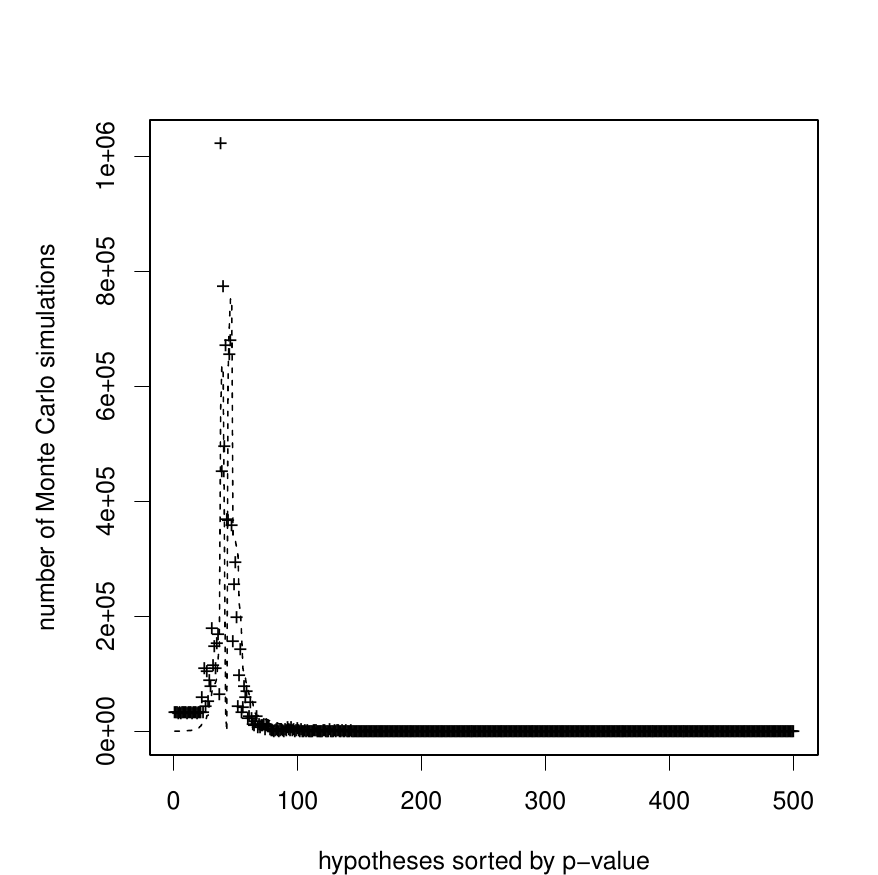}
\end{minipage}~
\begin{minipage}{0.32\textwidth}
\includegraphics[width=\textwidth]{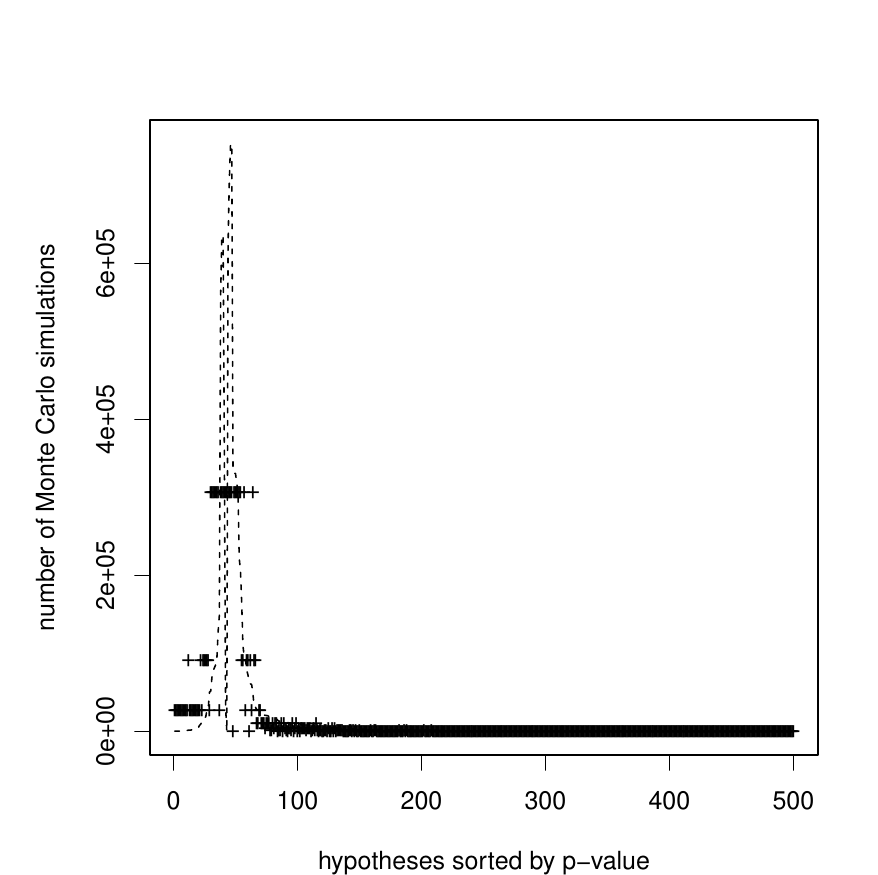}
\end{minipage}~
\begin{minipage}{0.32\textwidth}
\includegraphics[width=\textwidth]{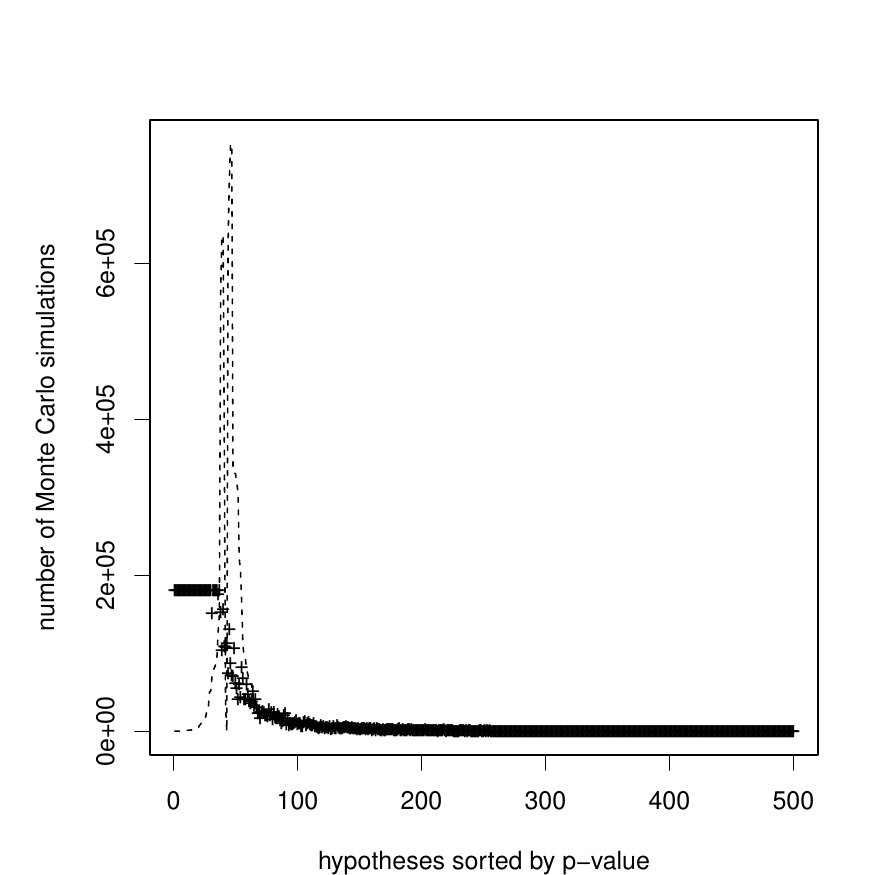}
\end{minipage}
\caption{Optimal KKT allocation computed as in Section~\ref{sec:without_pseudo} (dashed line) and integer allocation (crosses) of \texttt{QuickMMCTest} (left), \cite{GuoPedadda2008} (middle) and \cite{Sandve2011} (right). Subsampled dataset of \cite{Pekowska2010} with $m=500$. No pseudo-count ($c=0$ in \eqref{eq:derivh}).\label{fig:pekowska}}
\end{figure}

This section compares \texttt{QuickMMCTest} to other algorithms on a dataset of gene modifications (so-called \textit{H3K4me2} modifications) of \cite{Pekowska2010}. This dataset was used as a motivating example in the original publication of \texttt{QuickMMCTest} in \cite{GandyHahn2017}. The dataset contains midpoints of gene modifications on a genome, and the permutation test of \cite[Section~3.2]{Sandve2011} is used in \cite{GandyHahn2017} to test if gene modifications appear more often in the lower half of the genome. Preparing the dataset as outlined in \cite[Section~3]{GandyHahn2017} leads to $3465$ hypotheses under consideration.

\texttt{QuickMMCTest} is compared to the seven algorithms listed in Section~\ref{sec:introduction}, precisely the na\"ive method which generates a constant number of $K/m$ simulations per hypothesis as well as the algorithms of \cite{BesagClifford1991}, \cite{GuoPedadda2008}, \cite{Wieringen2008}, \cite{Sandve2011}, \cite{JiangSalzman2012}, and \cite{GandyHahn2014}. Each of those algorithms relies on one or more parameter, and the specific choice of parameters employed in this section is given in Section~\ref{sec:parameters} for each algorithm.

The optimal allocation derived in Section~\ref{sec:normalapprox} requires full knowledge of the p-values of all hypotheses, which are actually unavailable for a real dataset. Therefore, the p-values of all hypotheses are approximated once using $10^6$ permutations per hypothesis. The resulting p-value estimates (computed with a pseudo-count) are used to both compute the optimal KKT allocation of Monte Carlo simulations as well as to model the number of exceedances by drawing Binomial samples as described in Section~\ref{sec:formulation}. Moreover, computing the optimal allocation for a p-value distribution with many hypotheses and a high proportion of true nulls can be computationally challenging due to numerical instabilities of the KKT derivatives. Therefore, to simply computations, a subsample of size $m=500$ of the $3465$ p-values of the \cite{Pekowska2010} dataset is taken once without replacement, since such a subsample preserves the overall shape of the p-value distribution. All algorithms are applied to this subsample in a single run.

All algorithms were given $K=10^7$ simulations to allocate. Testing was carried out using a corrected Bonferroni threshold of $\alpha=0.1/m$.

Figure~\ref{fig:pekowska} shows the allocation of simulations for \texttt{QuickMMCTest} (left) as well as for \cite{GuoPedadda2008} (middle) and \cite{Sandve2011} (right). As observed in Figure~\ref{fig:QMT_and_KKT}, \texttt{QuickMMCTest} yields an allocation of a qualitative similar shape as the optimal KKT allocation. The algorithm of \cite{GuoPedadda2008} approximates the shape of the optimal KKT allocation by allocating large numbers of batches to the left and to the right of the KKT peak. Though not shown here, the algorithms of \cite{Wieringen2008} and \cite{GandyHahn2014} produce similar allocations. In contrast, the algorithm of \cite{Sandve2011} closely approximates the right half of the KKT allocation (corresponding to larger p-values), and uses an upper bound on the number of simulations that hypotheses with small p-values receive. The algorithms of \cite{BesagClifford1991} and \cite{JiangSalzman2012} produce similar allocations (figures not shown here). The na\"ive method distributes a constant number of Monte Carlo simulations to each hypothesis which results in the worst approximation of the KKT allocation.

\begin{table}
\centering
\begin{tabular}{l|l}
algorithm & $\Vert k-k_\text{KKT}^\ast \Vert/K$\\
\hline
\texttt{QuickMMCTest} & 0.115\\
na\"ive method & 0.196\\
\cite{BesagClifford1991} & 0.193\\
\cite{GuoPedadda2008} & 0.134\\
\cite{Wieringen2008} & 0.187\\
\cite{Sandve2011} & 0.191\\
\cite{JiangSalzman2012} & 0.174\\
\cite{GandyHahn2014} & 0.173
\end{tabular}
\caption{Normalised $L_2$ difference of the allocation vector $k$ of each method to the optimal KKT allocation $k_\text{KKT}^\ast$ for $K=10^7$.\label{tab:pekowska}}
\end{table}
Table~\ref{tab:pekowska} shows the normalised $L_2$ difference $\Vert k-k_\text{KKT}^\ast \Vert/K$ of the allocation vector $k$ of each algorithm to the optimal KKT allocation $k_\text{KKT}^\ast$ (see Section~\ref{sec:real_and_integer_allocation}). The table shows that indeed, the allocation of \texttt{QuickMMCTest} yields the closest allocation to the KKT one, followed by the algorithm of \cite{GuoPedadda2008}. Empirically it turns out that \texttt{QuickMMCTest} often allocates considerably more simulations than the optimal KKT allocation in the peak around the threshold (see Figure~\ref{fig:pekowska}, left), a fact which worsens the quality of its allocation, and that as $K$ increases, the discrepancy of \texttt{QuickMMCTest} to the other algorithms decreases.

\section{Discussion}
\label{sec:discussion}
This article considered the problem of allocating a fixed number of $K \in \N$ Monte Carlo simulations to $m \in \N$ hypotheses tested with the Bonferroni correction in order to approximate p-values. When estimating p-values both with or without a pseudo-count \citep{DavisonHinkley1997}, the optimal real-valued (and normally approximated) allocation is derived and computed, and a scheme based on simulated annealing is proposed to compute an approximation to the optimal integer allocation.

A simulation study shows that the real-valued and normally approximated optimal KKT solution is qualitatively similar to the SA integer solution, and moreover that the relative difference between both allocations seems to decrease (to zero) as $m,K,s^{\max} \rightarrow \infty$. Moreover, the allocation returned by the \texttt{QuickMMCTest} algorithm of \cite{GandyHahn2017} is compared to the optimal KKT solution. \texttt{QuickMMCTest} approximates unknown p-values at runtime while the testing of all hypotheses is in progress, and aims to efficiently allocate the $K$ Monte Carlo simulations to the hypotheses whose decision (rejected or non-rejected) is most ``unstable'' (see Section~\ref{sec:quickmmctest}). Simulations show that the allocation of \texttt{QuickMMCTest} computed at runtime seems to asympotically coincide with the optimal KKT solution, thus making \texttt{QuickMMCTest} a very attractive method to carry out multiple testing in practice. The results also give an intuition behind the low numbers of misclassifications already observed for this algorithm in \cite{GandyHahn2017}.

The current article leaves scope for further avenues of research. First, since a derivation of the optimal integer allocation seems infeasible (due to the fact that the problem is non-convex), it would be worth investigating how far away the optimal real-valued KKT solution is from (one of the) optima of the integer allocation: this could, in principle, be approached using subdifferential versions of the KKT conditions \citep{Ruszczynski2006}. Second, a theoretical analysis of \texttt{QuickMMCTest} could lead to an intuition for a formal proof that its allocation indeed satisfies some kind of asympotic optimality. This is not entirely unlikely since \texttt{QuickMMCTest} essentially borrows its strength from Thompson sampling \citep{Thompson1933}, for which optimality statements have already been proven in the related context of multi-armed bandit methodology \citep{AgrawalGoyal2012}.

\appendix
\section{Auxiliary lemma}
\label{sec:proof}
\begin{lemma}
\label{lemma:strictly}
If $c=0$ and $p_i \neq \alpha$, the derivatives $\partial h/\partial k_i$ defined in \eqref{eq:derivh} are negative and strictly increase to zero as $k_i \rightarrow \infty$ for all $i \in I$.
\end{lemma}
\begin{proof}
Case $p_i \leq \alpha$: Substitute $c=0$ into \eqref{eq:derivh} and write $\partial h_i/\partial k_i = - A(k_i) \cdot B(k_i)$, where
$$A(k_i) = \frac{k_i(\alpha-p_i)}{2 k_i \sqrt{k_i p_i (1-p_i)}} = \frac{\alpha-p_i}{2 \sqrt{k_i p_i (1-p_i)}},
\qquad B(k_i) = \phi \left( \frac{k_i(\alpha-p_i)}{\sqrt{k_i p_i (1-p_i)}} \right).$$
Since $p_i \geq 0$, $k_i \geq 0$ and $0 < \alpha-p_i$ (since $p_i \neq \alpha$), and as $\phi$ is positive, it follows that $A(k_i)>0$ and $B(k_i)>0$, thus $\partial h/\partial k_i < 0$.

As $A(k_i) \propto k_i^{-1/2}$ and $B(k_i) \propto \phi \left( \sqrt{k_i} \right)$, both functions $A$ and $B$ converge to zero as $k_i \rightarrow \infty$, thus $\partial h/\partial k_i \rightarrow 0$ as $k_i \rightarrow \infty$.

For $k_i < k_i'$, $A(k_i) > A(k_i')$. Likewise, $B(k_i) > B(k_i')$ for $k_i < k_i'$. Thus both functions $A$ and $B$ are strictly decreasing, and so is their product $A(k_i) \cdot B(k_i)$, implying that $\partial h_i/\partial k_i = - A(k_i) \cdot B(k_i)$ is strictly increasing.

The case $p_i > \alpha$ is proven similarly.
\end{proof}

Under the assumption that the p-values $p=(p_1,\ldots,p_m)$ are drawn from a distribution which is absolutely continuous with respect to the Lebesgue measure, and since $I=\{ 1,\ldots,m \}$ for a finite $m \in \N$, the probability of the event $\{ \exists i \in I: p_i = \alpha \}$ is zero. The condition $p_i \neq \alpha$ in Lemma~\ref{lemma:strictly} is therefore not a restriction when computing the optimal (real-valued) allocation for randomly drawn p-values.

\section{Repetition of the comparison with \texttt{QuickMMCTest}}
\label{sec:highpi}
\begin{figure}[t]
\centering
\begin{minipage}{0.49\textwidth}
\includegraphics[width=\textwidth]{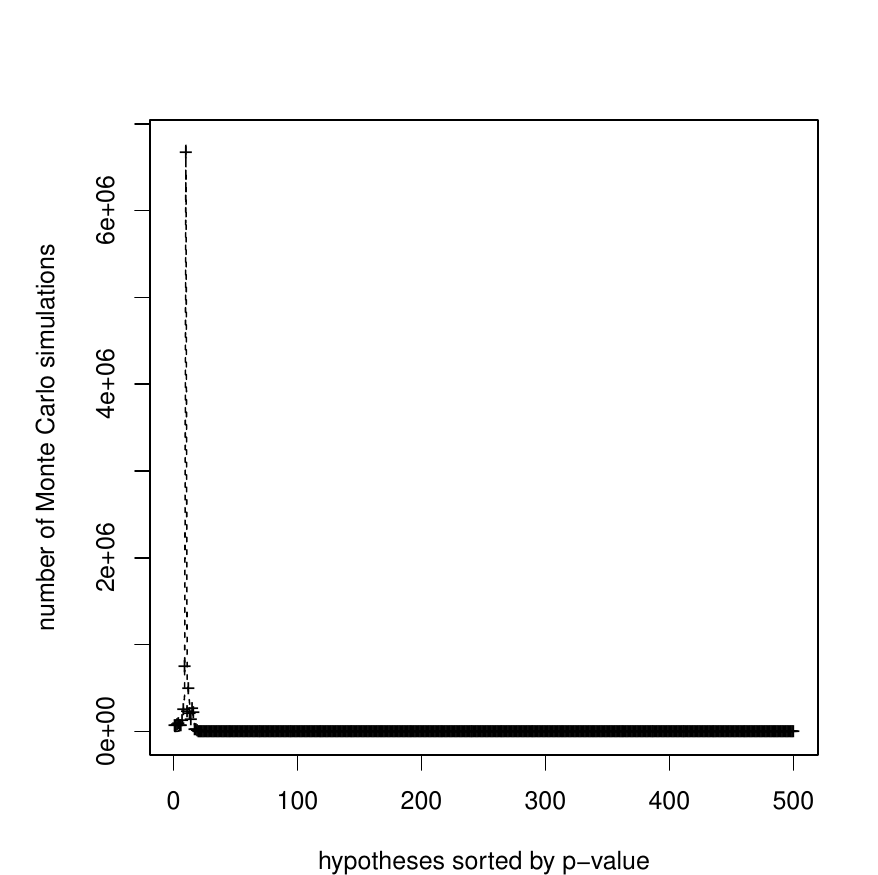}
\end{minipage}~
\begin{minipage}{0.49\textwidth}
\includegraphics[width=\textwidth]{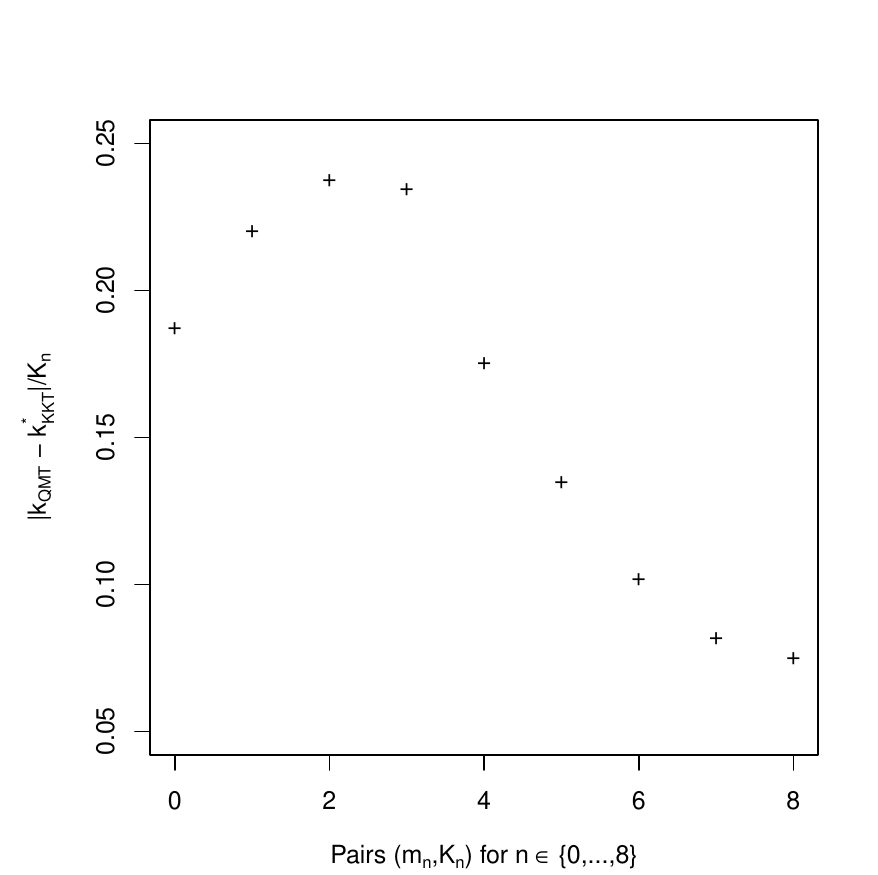}
\end{minipage}
\caption{Left: Optimal KKT allocation computed as in Section~\ref{sec:without_pseudo} (dashed line) and integer allocation of \texttt{QuickMMCTest} (crosses). No pseudo-count ($c=0$ in \eqref{eq:derivh}). Right: Relative difference in vectors $k_\text{QMT}$ (allocation of \texttt{QuickMMCTest}) and $k_\text{KKT}^\ast$ (optimal KKT allocation computed as in Section~\ref{sec:without_pseudo}) as a function of $(m_n,K_n)$, $n \in \{0,\ldots,8\}$.\label{fig:QMT_and_KKT_pi09}}
\end{figure}
Figure~\ref{fig:QMT_and_KKT_pi09} repeats the comparison of the optimal real-valued KKT allocation to the one of the \texttt{QuickMMCTest} algorithm for a dataset of $m=500$ p-values generated from the \cite{Sandve2011} distribution with $\pi=0.9$. A proportion of true null hypotheses close to one is what would be expected in real data studies.

As in Section~\ref{sec:quickmmctest}, the total number of simulations was $K=10^7$, a standard Bonferroni type threshold of $\alpha=0.1/m$ was employed, and \texttt{QuickMMCTest} was run with parameter $\Delta=10m$.

Figure~\ref{fig:QMT_and_KKT_pi09} (left) shows that \texttt{QuickMMCTest} again captures well the spike in the optimal allocation of Monte Carlo simulations. Figure~\ref{fig:QMT_and_KKT_pi09} (right) shows the relative difference $\Vert k_\text{QMT}-k_\text{KKT}^\ast \Vert/K_n$ in $L_2$ norm between the two allocation vectors, which is normalised with respect to the number of simulations $K_n$. As in Section~\ref{sec:quickmmctest}, the parameters $m$ and $K$ are increased together as $(m_n,K_n) = (50 \cdot 2^n, 10^5 \cdot 2^n)$ for $n \in \{0,\ldots,8\}$. Each datapoint is the median of $100$ repetitions. The figure shows that after an initial slight increase, the difference between the optimal KKT allocation and the one of \texttt{QuickMMCTest} decreases as $n$ increases. The origin of the initial slight increase is unknown and remains for further research.

\section{Choice of parameters for published methods}
\label{sec:parameters}
The algorithms employed in Section~\ref{sec:pekowska} were run with the following choice of parameters:
\begin{enumerate}
  \item The na\"ive method generated $K/m$ simulations per hypothesis.
  \item \cite{BesagClifford1991} sequentially generated one Monte Carlo simulation at a time for each hypothesis until either $h=20$ exceedances (as proposed by the authors) were observed (in which case this hypothesis was excluded from receiving further simulations) or the total number of simulations $K$ was reached.
  \item \cite{GuoPedadda2008} generated simulations according to a geometric sequence $B_0 \leq B_1 \leq \ldots \leq B_N$, where $N=9$ and the geometric increase was chosen such that $\sum_{i=0}^N B_i=K$. Confidence intervals of \cite{ClopperPearson1934} at a confidence level $0.01$ were employed as proposed by the authors.
  \item \cite{Wieringen2008} was employed using the $0.001$ upper quantile of the standard Normal distribution (as proposed by the authors) and a batch size of $100$ simulations.
  \item \texttt{MCFDR} of \cite{Sandve2011} was run with a \cite{BesagClifford1991} cutoff of $h=20$ exceedances and a batch size of $100$ for generating new simulations.
  \item \cite{JiangSalzman2012} was employed with parameters $a=10$, $\delta=0.01$, and a batch size of one (as proposed by the authors in their simulation study).
  \item \texttt{MMCTest} of \cite{GandyHahn2014} was run with \cite{Lai1976} confidence sequences at a confidence level of $0.01$, a batch size of $10$ simulations, and a geometric increase to match the total number of simulations $K$ over all $10$ iterations as done for \cite{GuoPedadda2008}.
\end{enumerate}


\end{document}